\documentclass[a4paper,11pt]{article}
\usepackage{jcappub} 
\usepackage{lineno,booktabs}
\usepackage{microtype}
\usepackage{fancyhdr} 
\usepackage{standalone} 
\usepackage{hyperref}
\usepackage{bold-extra}
\usepackage{graphicx}
\usepackage{amsmath}
\usepackage{amstext,fontawesome}
\usepackage{tablefootnote}
\usepackage{amssymb}
\usepackage{xspace}
\usepackage{comment}
\usepackage{wrapfig}
\usepackage{xcolor}
\usepackage{natbib}
\usepackage{amsmath,bm}

\definecolor{notes}{HTML}{C70039}
\definecolor{links}{HTML}{2D77A8}

\hypersetup{
    colorlinks=true,
    linkcolor=blue,
    filecolor=magenta,      
    urlcolor=links, 
}
\mathchardef\mhyphen="2D

\newlength{\dhatheight}

\newcommand{\biwei}[1]{{\textcolor{red}{Biwei: #1}}}

\newcommand{\kn}[1]{{\textcolor{blue}{KN: #1}}}

\title{Detecting Modeling Bias with Continuous Time Flow Models on Weak Lensing Maps}

\author[a,b]{Kangning Diao,}
\author[c]{Biwei Dai,}
\author[b,d,e]{and Uroš Seljak}
\affiliation[a]{Department of Astronomy, Tsinghua University, Beijing, 100084, China}
\affiliation[b]{Berkeley Center for Cosmological Physics, University of California, Berkeley, CA 94720, USA}
\affiliation[c]{School of Natural Sciences, Institute for Advanced Study, 1 Einstein Drive, Princeton, New Jersey 08540, USA}
\affiliation[d]{Physics Division, Lawrence Berkeley National Lab, 1 Cyclotron Road, Berkeley, CA 94720, USA}
\affiliation[e]{Department of Physics, University of California, Berkeley, CA 94720, USA}

\emailAdd{dkn20@mails.tsinghua.edu.cn}
\emailAdd{biwei@ias.edu}
\emailAdd{useljak@berkeley.edu}
\abstract{
Simulation-based inference  (SBI) provides a powerful framework for extracting rich information from nonlinear scales in current and upcoming cosmological surveys, and ensuring its robustness requires stringent validation of forward models.
In this work, we recast forward model validation as an out-of-distribution (OoD) detection problem within the framework of machine learning (ML)-based SBI. We employ probability density as the metric for OoD detection, and compare various density estimation techniques, demonstrating that field-level probability density estimation via continuous time flow models (CTFM) significantly outperforms feature-level approaches that combine scattering transform (ST) or convolutional neural networks (CNN) with normalizing flows (NFs), as well as NF-based field-level estimators, as quantified by the area under the receiver operating characteristic curve (AUROC). Our analysis shows that CTFM not only excels in detecting OoD samples but also provides a robust metric for model selection. Additionally, we verified CTFM maintains consistent efficacy across different cosmologies while mitigating the inductive biases inherent in NF architectures. Although our proof-of-concept study employs simplified forward modeling and noise settings, our framework establishes a promising pathway for identifying 
unknown systematics in the
cosmology datasets.
}

\begin{document}
\maketitle
\flushbottom

\section{Introduction}

Cosmology has entered an era of precision science, with current models predicting a wide range of observables at sub-percent level accuracy. Concurrently, the next generation of telescopes promises to deliver unprecedented volumes of high-quality, multi-observable data. Among these, weak gravitational lensing (WL) \citep{2001PhR...340..291B,2015RPPh...78h6901K}, which is the subtle distortion of light from distant galaxies induced by intervening large-scale structures, stands out as a key probe for mapping the total matter distribution in the universe \citep[e.g.][]{2017MNRAS.465.1454H,2017MNRAS.465.2033J,2019PASJ...71...43H,2020PASJ...72...16H}. Upcoming surveys by facilities such as \href{https://RubinObservatory.org}{LSST} \citep{2019ApJ...873..111I}, \href{https://www.esa.int/Science_Exploration/Space_Science/Euclid}{Euclid} \citep{2022A&A...662A.112E}, and \href{https://roman.gsfc.nasa.gov}{Roman} \citep{2015arXiv150303757S} are expected to revolutionize our understanding of cosmic origins, composition, and evolution. 

A variety of summary statistics have been developed to analyze WL data, beginning with the traditional N-point correlation functions \citep{2017MNRAS.469.2737K,2003MNRAS.344..857T,2003ApJ...584..559Z,2014MNRAS.441.2725F,2011MNRAS.410..143S}. However, these methods are often hampered by issues such as incomplete information capture \citep{2011ApJ...738...86C}, the proliferation of statistical coefficients, large variances, and a higher sensitivity to outliers. To address these issues, researchers have proposed alternative approaches including correlation functions computed on transformed or marked fields \citep{2009ApJ...698L..90N,2016JCAP...11..057W}, peak counts \citep{2000ApJ...530L...1J,2010PhRvD..81d3519K}, void statistics \citep{2019BAAS...51c..40P}, Minkowski functionals \citep{1994A&A...288..697M,2012PhRvD..85j3513K}, scattering transforms (ST) \citep{Mallat_2012,Cheng_2021,Cheng_2020,2022PhRvD.105j3534V,2020PhRvD.102j3506A}, and features extracted via convolutional neural networks (CNN) and other neural network architectures \citep{2018PhRvD..97j3515G,2021JCAP...11..049M,2022MNRAS.511.1518L,2023MNRAS.521.2050L,2018PhRvD..97h3004C,2024arXiv241007548M}. Typically, the likelihoods associated with these summary statistics are modeled using either multivariate Gaussian approximations or simulation-based inference (SBI), yet these approaches remain susceptible to their ad hoc nature and potential information loss. Recently, the advent of advanced machine learning models and increased computational power has led to the development of normalizing flows for field-level likelihood modeling \citep[e.g.][]{2022MNRAS.516.2363D,2022ApJ...937...83H}, offering significant improvements over traditional feature-level methods.

Despite these advances, a critical challenge persists: the robustness of the forward models that underpin the training data. Variations among different hydrodynamical simulations and baryon models, none of which have converged to a single, universally accepted description \citep[e.g.][]{2019MNRAS.488.1652H}, can introduce biases. Models trained on one simulation may perform poorly when applied to data from another \citep{2021arXiv210909747V}, and there is no guarantee that any current forward model accurately represents the real universe. It is therefore imperative to detect whether a forward model deviates from actual observations. Questions naturally arise, such as which features are reliable, and which may be compromised by unmodeled or inaccurately modeled effects? How can we identify these discrepancies in high-dimensional, complex data? In the context of field-level inference, these issues are particularly important. The richer the information incorporated into the inference pipeline, the greater the risk of inadvertently including features that are poorly modeled, potentially leading to biased posteriors \citep[e.g. Figure 5 in][]{2024PNAS..12109624D} or overconfident constraints.

Beyond their impact on inference, these biases also signal gaps in our understanding of the underlying physics. Discrepancies between forward models and observations may arise from diverse sources—including cosmological evolution, complex astrophysical processes, and unaccounted observational effects, and thus merit thorough investigation. Detecting such biases can be framed as a consistency test: given that the full range of forward model outputs forms a statistical distribution, the task becomes one of determining whether an observation is a member of that distribution, which is often called the out-of-distribution (OoD) detection. When the likelihood $\mathcal{L}$ is Gaussian, $-2\log \mathcal{L}$ follows the $\chi^2$-distribution and is widely used to assess how well the model fits the data in current survey analysis \citep[e.g.][]{asgari2021kids,secco2022dark,li2023hyper}. Recently, \cite{2024PNAS..12109624D} generalizes the test to high-dimensional field-level inference where the likelihood is no longer Gaussian. They employed wavelet decomposition to segregate information by scale, leveraging the relative robustness of large-scale structures in modeling, and used normalizing flows to learn the corresponding distributions at each scale. Moreover, continuous time flow models (CTFM), such as diffusion models \citep{S-Dickstein2015,Song2021,Ho2020} and flow matching (FM) models \citep{Lipman2022}, have emerged as state-of-the-art techniques for learning high-dimensional distributions. These methods have seen extensive application in {sampling} various cosmological fields {across different tasks, including emulation \citep{2023MNRAS.526.1699Z}, super-resolution \citep{2024OJAp....7E.104S} and reconstruction of cosmological fields\citep{2025arXiv250204158B,2024arXiv240717667B}, but they are not yet widely used for computing the probability at the field level. \citep{2025ApJ...978...64M} estimates the lower bound of field-level probability with CTFM, while in this work we compute the field-level probability directly through integral of CTFM trajectory}. We explore the use of CTFM {probability density} for bias detection at the field level and compare its performance against normalizing flows applied at both the field and feature levels.

This paper is organized as follows. In Section \ref{sec:method} we detail our detection methodology, while Section \ref{sec:res} presents our primary results. We validate our field-level density estimation method on Gaussian random field in Section \ref{sec:grf}, and test the performance of different density estimation methods in Section \ref{sec:bcm}. The consistency of our results across different cosmological models is verified in Section \ref{sec:consis},  and Section \ref{sec:sel} discusses the application of out-of-distribution detection metrics for model selection, and Section \ref{sec:scale} examines the scalability of our approach to large survey volumes. 
We further analyze the challenges posed by the inductive bias inherent in NF models in Section \ref{sec:badnf}. Finally, Section \ref{sec:con} summarizes our conclusions. We provides detailed model architectures in Appendix \ref{app:model}.

\section{Methods and Datasets}
\label{sec:method}
\subsection{Out-of-Distribution detection as a consistency test}

\begin{figure}
    \centering
    \includegraphics[width=0.8\linewidth]{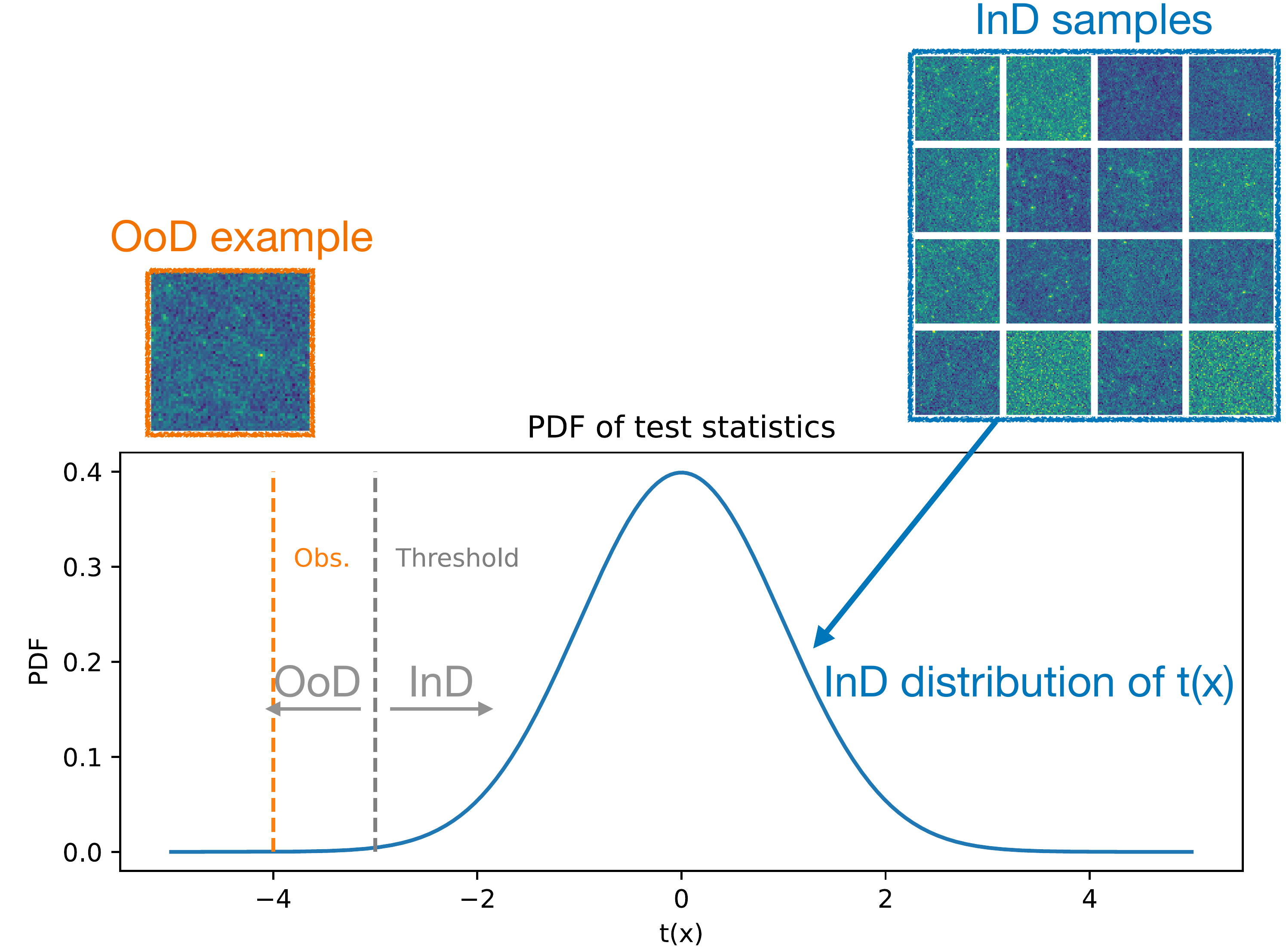}
    \caption{An illustration of OoD detection pipeline. A probability distribution function (PDF) of InD sample test statistics is calculated and shown in blue solid line, and an empirical threshold in grey dashed line is chosen to identify an OoD sample with $t(\bm x)$ smaller than the threshold, shown in orange dashed line. }
    \label{fig:detect}
\end{figure}
Detecting biases in forward modeling can be naturally framed as a consistency test between the model and the observation. When the forward model produces a full distribution of possible outputs, bias detection reduces to an OoD test, determining whether the observation is consistent with the model-generated distribution.

Following the posterior predictive test formalism \citep{gelman1996posterior}, we introduce an arbitrary test statistic \(t(\bm{x})\), where \(\bm{x}\) represents either simulated or observed data. Given an observation \(\bm{x}_{\rm obs}\) and its inferred posterior distribution \(p(\theta|\bm{x}_{\rm obs})\), the posterior predictive distribution is defined as
\begin{equation}
\label{eqn:ppd}
    p(\bm{x}_{\rm rep}|\bm{x}_{\rm obs}) = \int p(\bm{x}_{\rm rep}|\theta)p(\theta|\bm{x}_{\rm obs})\,d\theta,
\end{equation}
where \(\bm{x}_{\rm rep}\) denotes a replication of the observation. This distribution encapsulates all plausible outputs of the forward model, and samples drawn from it are considered in-distribution (InD).

If the model accurately captures the data, the value of the test statistic \(t(\bm{x})\) computed for the observation should be consistent with the distribution of \(t(\bm{x})\) values computed for replicated datasets. To quantify this consistency, we calculate the probability
\begin{equation}
    p\bigl(t(\bm{x}_{\rm rep})>t(\bm{x}_{\rm obs})\bigr) = \int \mathbf{1}_{\{t(\bm{x}_{\rm rep})>t(\bm{x}_{\rm obs})\}}\, p(\bm{x}_{\rm rep}|\bm{x}_{\rm obs})\,d\bm{x}_{\rm rep},
\end{equation}
where \(\mathbf{1}_A(\bm{x})\) is the indicator function, which equals 1 if \(\bm{x}\in A\) and 0 otherwise.

In practice, we draw \(N\) samples \(\{\bm{x}_{{\rm rep},i}\}_{i=1}^N\) from the posterior predictive distribution and compute the corresponding set of test statistics \(\{t(\bm{x}_{{\rm rep},i})\}_{i=1}^N\). If \(n\) out of these \(N\) samples satisfy \(t(\bm{x}_{\rm rep}) > t(\bm{x}_{\rm obs})\), then
\begin{equation}
    p\bigl(t(\bm{x}_{\rm rep})>t(\bm{x}_{\rm obs})\bigr)\approx \frac{n}{N}.
\end{equation}

Values of \(p(t(\bm{x}_{\rm rep})>t(\bm{x}_{\rm obs}))\) approaching 0 or 1 indicate that \(\bm{x}_{\rm obs}\) is likely an OoD sample relative to \(p(\bm{x}_{\rm rep}|\bm{x}_{\rm obs})\), suggesting potential biases in the forward model. To operationalize this test, an empirical threshold \(t_{\rm th}\) is often introduced, as is illustrated in Figure \ref{fig:detect}. For example, if OoD samples are flagged when \(t(\bm{x}_{\rm obs}) > t_{\rm th}\), the false positive rate (FPR) is estimated as \(p(t(\bm{x}_{\rm rep}) > t_{\rm th})\); similarly, if a lower threshold is used, the FPR is given by \(p(t(\bm{x}_{\rm rep}) < t_{\rm th})\). In cases where both upper and lower bounds are applied, the FPR is approximated as
\(
p(t(\bm{x}_{\rm rep}) < t_{\rm th,\ low}) + p(t(\bm{x}_{\rm rep}) > t_{\rm th,\ high}).
\) The choice of the bound depends on the choice of $t(\bm{x})$.

This framework provides a rigorous means of assessing whether an observation is consistent with the predicted distribution of outputs, thereby serving as a diagnostic for biases in the forward modeling process.

\subsection{Probability density as the test statistics}

Although the test statistic \(t(\bm{x})\) can be defined arbitrarily, selecting an informative \(t(\bm{x})\) significantly enhances the performance of detecting an anomaly. In the absence of a specific model for the anomaly there is no optimal choice for the test statistic. 
When searching for unknown unknowns 
we must therefore rely on test 
statistics that are generic. 
In this context, probability density estimation $p(\bm{x})$ is the most direct and intuitive choice, as values lower than the typical range for InD samples clearly indicate a low likelihood of sampling that particular observation from the modeled distribution. In
the Gaussian likelihood setup, the
density estimation is closely
related to the $\chi^2$ goodness-of-fit
test, $\ln p(\bm{x})=-\chi^2/2$, which is a widely used test for the 
validity of the model: a large value 
of $\chi^2$ compared to the number 
of degrees of freedom indicates that 
the model is a poor fit to the data, 
and thus a model misspecification that needs to be addressed. In this paper 
we generalize this concept to 
the non-Gaussian likelihoods learned
using ML. 

However, evaluating probability density in high-dimensional spaces is a non-trivial task. Here, we introduce several approaches to address this challenge. These approaches can be broadly categorized into two groups. The first involves feature-level methods, in which a compressor reduces the high-dimensional sample to a low-dimensional feature vector; the probability density of this vector is then estimated using an NF trained on the forward-modeled dataset. The second group comprises field-level density estimation techniques, where we directly evaluate the density of the high-dimensional observable using two variants of CTFM, the diffusion model and the optimal transport flow matching (OTFM) model. The detailed structures of different kinds of neural networks and corresponding training configurations mentioned from here on are described in Appendix \ref{app:model}.

\subsubsection{Feature-level density estimation}
\label{sec:feature}
In the feature-level density estimation, the high-dimensional field is first compressed into a low-dimensional feature vector, after which its probability density is estimated using a NF. In this work, we employ two compressors: the Scattering Transform (ST) coefficients and a CNN trained with the VMIM loss \citep{VMIM}. {Specifically, the CNN-learned statistic has been shown to provide optimal performance in constraining cosmological parameters on Gaussian random fields and log-normal fields \cite{lucas2021lossless,lanzieri2025optimal}, while on non-linear weak lensing maps, it also achieves similar performance to field-level analysis \cite{2024JCAP...08..010S}}.
After the compression, the real-valued non-volume preserving transformations (RealNVP) \citep{2016arXiv160508803D} is then used as the NF density estimator.

\paragraph{ST Compressor}
The ST coefficients, \(S_1\) and \(S_2\), are defined as
\begin{equation}
    \begin{aligned}
        \bm I_{1}(j,l) &= \left|\bm{x}\ast\Psi\left(j,l\right)\right|\ast\Phi(j),\\[1mm]
        \bm I_{2}(j_1,l_1,j_2,l_2) &= \left|\left|\bm{x}\ast\Psi\left(j_1,l_1\right)\right|\ast\Psi\left(j_2,l_2\right)\right|\ast\Phi(j_2),\\[1mm]
        S_{1}(j,l)& =  \left<\bm I_{1}(j,l)\right>,\\[1mm]
        S_{2}(j_1,l_1,j_2,l_2)& = \left<\bm I_{2}(j_1,l_1,j_2,l_2)\right>.
    \end{aligned}
\end{equation}
Here, \(\bm{x}\) is the input field, $\ast$ denotes the convolution operation, \(\Psi\) represents the Morlet wavelet kernel (see e.g. Appendix~B of \cite{Cheng_2020} for details) and $\Phi$ is the Gaussian kernel to filter all small-scale fluctuations. The index \(j\) specifies the scale of the convolutional kernel with smaller \(j\) corresponding to more localized kernels while \(l\) defines its orientation. We select \(j=0\text{–}3\) and \(l=0\text{–}3\) to cover a broad range of scales and orientations, yielding 16 coefficients for \(S_{1}(j,l)\) and 96 coefficients for \(S_{2}(j_1,l_1,j_2,l_2)\), for a total feature vector length of 112. The ST coefficients are computed using \textsc{Kymatio}\footnote{\url{https://github.com/kymatio/kymatio}} \citep{Andreux_2020}.

\paragraph{CNN Compressor}
For the CNN-based compressor, we utilize a 34-layer ResNet \citep{2016cvpr.confE...1H} optimized with the VMIM loss:
\begin{equation}
    (\bm w^*,\bm u^*) = \arg\min_{\bm w ,\bm u} \mathbb{E}_{p(\theta,\bm{x})}[-\log p_{\bm u}(\theta|f_{\bm w} (\bm{x}))],
\end{equation}
where an auxiliary RealNVP with parameter $u$ is used to estimate \(\log p_{\bm u}(\theta|f(\bm{x}))\), \(f_{\bm w}\) denotes the CNN with parameter $\bm w$, and \(f_{\bm w}(\bm{x})\) is the compressed feature vector. The dimension of this feature vector is set to 128 to align with the ST feature vector, and we also performed experiments with different feature dimensionality and verify that this length has a negligible effect in the OoD detection performance.

\

Once the ST and CNN compressors have been applied, separate RealNVPs are trained on the resulting feature vectors to estimate their probability densities. The NF loss function is defined as
\begin{equation}
    \bm w^* = \arg\min_{\bm w}\mathbb{E}_{p(\theta,\bm{y})}[-\log p_{\bm w}(\bm{y}|\theta)],
\end{equation}
with \(\bm{y}\) representing the compressed feature vector and $\bm w$ is the parameters of the NF. 

\subsubsection{Field-level density estimation}
\label{sec:field}
Unlike feature-level methods, field-level density estimation avoids the information loss inherent in compression, and is therefore expected to perform better. In this work, we employ two variants of CTFM, a diffusion model and an OTFM model, as field-level density estimators.

Given any distribution \(p(\bm{x})\), CTFM 
estimates the probability density of a sample \(\bm{x}\) by solving an ordinary differential equation (ODE) \cite{chen2018neural}. Specifically, such an ODE has the form
\begin{equation}
    \frac{d}{dt}\phi_t(\bm{x}) = f(\phi_t(\bm{x}),t),
\end{equation}
where the transformation \(\phi:[0,1]\times\mathbb{R}^d\rightarrow\mathbb{R}^d\) is termed the \textit{flow} and is initialized as \(\phi_0(\bm{x}) = \bm{x}\). As the time \(t\) varies from 0 to 1, the flow transports the original distribution \(p(\bm{x})\) to a tractable target distribution \(p(\phi_1(\bm{x}))\), typically a \(d\)-dimensional standard Gaussian. For simplicity, we denote \(\phi_t(\bm{x})\) as \(\bm{x}_t\). Assuming that the density \(p_1(\bm{x}_1)\) is tractable, the density \(p(\bm{x})\) is given by
\begin{equation}
    \log p(\bm{x}) = \log p_1(\bm{x}_1) + \int_0^1 \nabla\cdot f(\bm{x}_t,t)\, dt.
    \label{eqn:int}
\end{equation}
The divergence \(\nabla\cdot f(\bm{x}_t,t)\) is estimated using the Skilling-Hutchinson trace estimator \cite{Skilling1989,hutchinson1989stochastic}:
\begin{equation}
    \nabla\cdot f(\bm{x}_t,t)\approx\mathbb{E}_{\epsilon\sim p(\bm\epsilon)}\left[\bm\epsilon^T \nabla f(\bm{x}_t,t)\bm\epsilon\right],
    \label{eqn:sk estimator}
\end{equation}
where \(p(\bm\epsilon)\) is a distribution with zero mean and an identity covariance matrix. When \(f(\bm{x}_t,t)\) is implemented as a differentiable function, the term \(\bm\epsilon^T\nabla f(\bm{x}_t,t)\) can be computed via automatic differentiation of \(\bm\epsilon^T f(\bm{x}_t,t)\). In our experiments, we adopt a standard Gaussian for \(p(\bm\epsilon)\) and perform Euler integration with 1000 steps for Equation \ref{eqn:int}.

\paragraph{Diffusion models}
In diffusion models \citep{S-Dickstein2015,Ho2020}, the corresponding ODE is derived from a diffusion process described by the stochastic differential equation (SDE) \citep{Song2021}
\begin{equation}
    {d}\bm{x}_t  = -\frac{1}{2}\beta_t\bm{x}_t\, dt + \sqrt{\beta_t}\bm\, d\bm{b},
    \label{eqn:sde}
\end{equation}
where $\bm{b}$ is the Brownian motion and  \(\beta_t\) is a predefined, monotonically increasing function satisfying \(\beta_0=0\) and \(\beta_1\rightarrow +\infty\). This SDE gradually transforms \(p(\bm{x})\) into a standard Gaussian by incrementally adding noise and diminishing the influence of the initial sample. The associated probability flow ODE preserves the marginal density \(p_t(\bm{x}_t)\) for all \(t\) \citep{Song2021} and thus fulfills the requirement for transforming \(p(\bm{x})\) into a standard Gaussian, enabling the use of Equation \ref{eqn:int} to recover \(p(\bm{x})\). This probability flow ODE is given by \citep{Song2021,SongJ2020}
\begin{equation}
    d \bm{x}_t = -\frac{1}{2}\left(\beta_t\bm{x}_t  +\beta_t\nabla_{\bm{x}_t}\log p_t(\bm{x}_t)\right) dt.
    \label{eqn:pfode}
\end{equation}
Integration of this ODE requires knowledge of the score function \(\nabla_{\bm{x}_t}\log p_t(\bm{x}_t)\), which we approximate using a neural network \(\bm{s}_{\bm{w}}(\bm{x},t)\) parameterized by weights \(\bm{w}\). The network is trained by minimizing the loss \cite{Ho2020}
\begin{equation}
\begin{aligned}
    \bm{w}^* = \arg\min_{\bm{w}}\, \mathbb{E}_t\mathbb{E}_{\bm{x}_0\sim p_0(\bm{x})} \mathbb{E}_{{\bm{x}}_t\sim p_t({\bm{x}} \vert \bm{x}_0)}\\
    \left[\Vert \bm{s}_{\bm{w}}(\bm{x}_t,t)- \nabla_{\bm{x}_t}\log p_t(\bm{x}_t\vert \bm{x}_0)\Vert^2 \right].
\end{aligned}
\end{equation}
Given that the noise \(\bm\epsilon\) in Equation \ref{eqn:sde} is Gaussian, we have \citep{sarkka2019applied}
\begin{equation}
    \nabla_{\bm{x}_t}\log p_t(\bm{x}_t\vert \bm{x}_0) = \frac{\sqrt{\alpha_t} \bm{x}_0 -\bm{x}_t}{1-\alpha_t},
\end{equation}
with \(\alpha_t:=\exp\left(-\frac{1}{2}\int_0^t\beta_s\,ds\right)\). Since the probability flow ODE and the SDE share the same marginal distribution \(p_t(\bm{x}_t)\), they also share the same score function. By substituting the neural network \(\bm{s}_{\bm{w}}(\bm{x},t)\) in place of the true score in Equation \ref{eqn:pfode}, we obtain the diffusion model ODE for estimating \(p(\bm{x})\). Our implementation of the diffusion model, including the neural network, is based on the \texttt{diffusers}\footnote{\url{https://huggingface.co/docs/diffusers/index}} package. We choose the linear schedule $\beta_t = 20t$ in this work.

\paragraph{Optimal Transport Flow Matching}
The diffusion model ODE defined in Equation \ref{eqn:pfode} requires the neural network to predict a non-constant score term over different $t$, which can pose challenges due to the complexity of the output. Optimal transport (OT) \citep{mccann1997convexity} addresses this issue by assuming a constant vector field \(f(\bm{x}_t,t) = \bm{x}_1-\bm{x}_0\), where \(p(\bm{x}_1)\) is a high-dimensional standard Gaussian. This simplification improves fitting \(f(\bm{x}_t,t)\) with a neural network. However, directly training under this assumption is challenging without explicit \((\bm{x}_0,\bm{x}_1)\) pairs. FM \citep{Lipman2022} overcomes this limitation by learning the OT vector field \(f(\bm{x}_t,t)\) without requiring explicit pairings.

In FM, given a sample \(\bm{x}_0\) drawn from \(p(\bm{x})\), the conditional distribution \(p_t(\bm{x}_t|\bm{x}_0)\) is modeled as a Gaussian with mean \(\bm{\mu}_t\) and standard deviation \(\bm{\sigma}_t\), i.e.,
\[
p_t(\bm{x}_t|\bm{x}_0) = \mathcal{N}(\bm{\mu}_t,\bm{\sigma}_t),
\]
and the marginal density is obtained by
\begin{equation}
    p_t(\bm{x}_t) = \int p_t(\bm{x}_t|\bm{x}_0)p(\bm{x}_0) \, d\bm{x}_0.
\end{equation}
By choosing $\bm{\mu}_0 = \bm{x}_0$, $\bm{\sigma}_0 = 0$, $\bm{\mu}_1 = 0$, and $\bm{\sigma}_1 = \bm{1}$, we ensure that $p_0(\bm{x}_0)=p(\bm{x})$ and $p_1(\bm{x}_1)$ is a standard Gaussian. Although an analytical expression for $f(\bm{x}_t,t)$ under this setting is difficult to obtain, it can be learned by training a neural network $\bm{s}_{\bm{w}}(\bm{x},t)$ through minimizing the loss \citep{Lipman2022}
\begin{equation}
\begin{aligned}
    \bm{w}^* = \arg\min_{\bm{w}}\, \mathbb{E}_{t}\mathbb{E}_{\bm{x}_0\sim p_0(\bm{x})}\mathbb{E}_{\bm{x}_t \sim p_t(\bm{x}_t|\bm{x}_0)}\\
    \left[\Vert\bm{s}_{\bm{w}}(\bm{x}_t,t) - \frac{d{\bm\sigma}_t(\bm{x}_0)}{{\bm{\sigma}}_t(\bm{x}_0)dt}(\bm{x}_t - \bm{\mu}_t) -\frac{d\bm{\mu}_t(\bm{x}_0)}{dt}\Vert^2\right].
    \label{eqn:cfm}
\end{aligned}
\end{equation}
To align this loss with the OT ODE, one can set \(\bm{\mu}_t = (1-t)\bm{x}_0\) and \(\bm{\sigma}_t = t\), so that \(\bm{x}_t = (1-t)\bm{x}_0 + t\bm\epsilon\) with \(\bm\epsilon\sim\mathcal{N}(0,\bm{1})\). Under these choices, Equation \ref{eqn:cfm} reduces to
\begin{equation}
    \begin{aligned}
    \bm{w}^* = \arg\min_{\bm{w}}\, \mathbb{E}_{t}\mathbb{E}_{\bm{x}_0\sim p_0(\bm{x})}\mathbb{E}_{\bm{\epsilon} \sim \mathcal{N}(0,\bm{1})}\\
    \left[\Vert\bm{s}_{\bm{w}}(\bm{x}_t,t) - (\bm{\epsilon} - \bm{x}_0)\Vert^2\right].
\end{aligned}
\end{equation}
Since \(\bm{\epsilon}\) and \(\bm{x}_1\) follow the same distribution, this loss function effectively trains \(\bm{s}_{\bm{w}}(\bm{x}_t,t)\) to approximate the difference \(\bm{x}_1 - \bm{x}_0\). Our implementation of OTFM is based on the \texttt{torchcfm} package \citep{2023arXiv230200482T}\footnote{\url{https://github.com/atong01/conditional-flow-matching}}.

\subsection{Weak lensing maps}
\subsubsection{Dark matter only maps}

The dark-matter-only (DMO) weak lensing (WL) convergence maps used in this study are obtained from \cite{2018PhRvD..97j3515G} and are generated from a suite of 80 N-body simulations under flat \(\Lambda\)CDM cosmologies. In these simulations, the baryon density, Hubble parameter, and spectral index are fixed at \(\Omega_b = 0.046\), \(h=0.72\), and \(n_s = 0.96\), respectively, while \(\Omega_m\) and \(\sigma_8\) vary around \(\Omega_m = 0.26\) and \(\sigma_8 = 0.8\).

The N-body simulations employ \(512^3\) particles in a box of size \(240h^{-1}\) Mpc and are run using GADGET-2 \citep{2005MNRAS.364.1105S}. WL convergence maps are computed via ray-tracing using a multiple-lens-plane algorithm \cite{1992grle.book.....S} on snapshots spanning \(0 < z < 1\). From each snapshot, an \(80h^{-1}\) Mpc slice along the line-of-sight is extracted as a lens plane. By applying random rotations, flips, and shifts to the snapshots, 512 pseudo-independent maps are derived from each simulation. Further details on the map generation process can be found in \cite{2018PhRvD..97j3515G}.

The convergence maps are produced at a resolution of \(1024\times1024\) pixels and are subsequently downscaled to various resolutions via spatial averaging. The noise considered in this work is galaxy shape noise, modeled as independent Gaussian noise in each pixel with a standard deviation given by
\begin{equation}
    \sigma_g = \frac{\sigma_\epsilon}{\sqrt{2n_gA_{\rm pix}}},
\end{equation}
where \(\sigma_\epsilon\sim 0.4\) denotes the mean intrinsic ellipticity of galaxies, \(n_g\) is the galaxy number density, and \(A_{\rm pix}\) is the pixel area. We investigate three noise scenarios corresponding to \(n_g = \{30,50,100\}\ {\rm arcmin}^{-2}\). The \(30\ {\rm arcmin}^{-2}\) case represents the targeted noise level for surveys such as LSST or Euclid, while future space missions like Roman may achieve \(50\ {\rm arcmin}^{-2}\) or higher. The \(100\ {\rm arcmin}^{-2}\) scenario is an optimistic forecast for forthcoming space-based surveys.

\subsubsection{Baryon effects}

To better model the physical process, baryon models are used to post-process the N-body simulation output \citep{2022MNRAS.511.1518L,2021MNRAS.506.3406L}. The post-processing first identify all the halos with mass larger than $10^{12}M_{\odot}$, then substitute the halo particles with spherical symmetric analytical density profile. The analytical halo profile derived from the Baryon Correction Model (BCM) \citep{2020MNRAS.495.4800A} represents halos as composed of four components: the central galaxy, bound gas, ejected gas due to AGN feedback, and relaxed dark matter. It is characterized by four free parameters: \(M_c\) (the halo mass that retains half of the total gas), \(M_{1,0}\) (the halo mass corresponding to a galaxy mass fraction of 0.023), \(\eta\) (the maximum distance to which gas is ejected from its parent halo), and \(\beta\) (the logarithmic slope that describes how the gas fraction scales with halo mass). Although this model omits the substructures and non-spherical shapes of halos, it has been argued that the morphological differences between simulated halos and these idealized spherical profiles are statistically negligible relative to the uncertainties in the power spectrum and peak counts measured in an HSC-like survey \citep{2021MNRAS.506.3406L}. 

These post-processed snapshots go through the same pipeline to produce the BCM convergence map with the same resolution and noise. For each cosmology, 2048 BCM maps are generated, each with distinct baryon parameter $\{M_c,M_{1,0},\eta,\beta\}$.

\section{Results}

\label{sec:res}

\subsection{Testing the field-level density estimator on Gaussian random fields}
\label{sec:grf}
\begin{figure}
    \centering
    \includegraphics[width=0.8\linewidth]{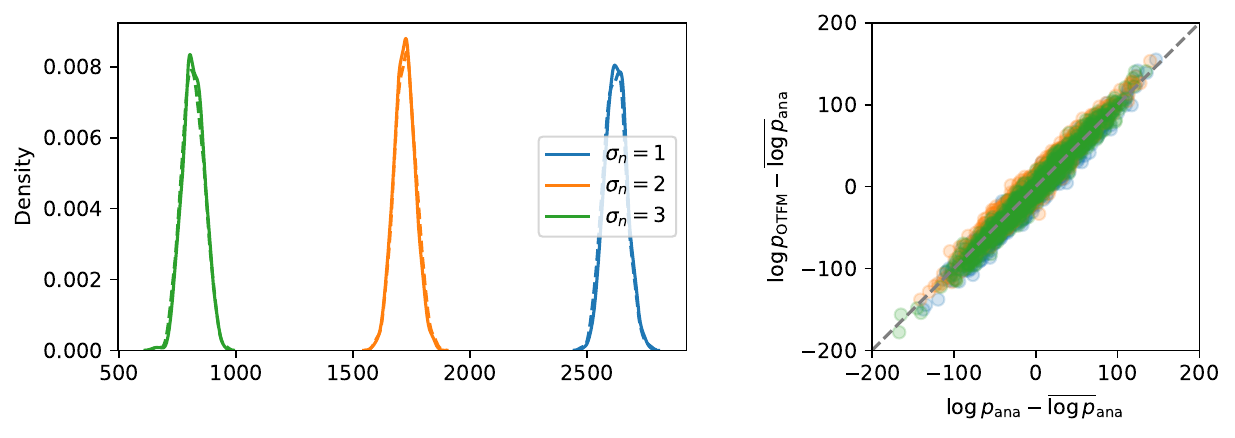}
    \caption{\textbf{Left}: Distribution of the log–probability densities of Gaussian random fields (GRFs) evaluated with OTFM (dashed curves) and by the analytic expression (solid curves). Different colours correspond to different noise levels~$\sigma_n$.  
    \textbf{Right}: Point‑by‑point comparison between the OTFM log density and the analytic value.  For each noise level, the mean log density inferred analytically has been subtracted, so that only the residual variations are shown.}
    \label{fig:grf}
\end{figure}

Gaussian random fields (GRFs) provide an ideal benchmark {for probability density estimation methods} because their log–likelihood can be computed in closed form.  Therefore, we first validates our density estimation formalism on GRFs.
We generate GRFs whose power spectrum is a power law with additive white noise,  
\[
P(k)=\bigl(k/5\bigr)^{-2}+\sigma_n,
\]
and test three noise amplitudes, $\sigma_n\in\{1,2,3\}$.  
For each~$\sigma_n$, we draw $120{,}000$ realisations of size~$64^2$ pixels and train an OTFM model; its performance is assessed on an independent set of $1024$ samples, yielding the OTFM estimated~$\log p_{\mathrm{OTFM}}$ {by applying OTFM to Equation \ref{eqn:int}}.  

The left panel of Figure~\ref{fig:grf} compares the distributions of $\log p_{\mathrm{OTFM}}$ with the analytic log density~$\log p_{\mathrm{ana}}$.  
The OTFM successfully recovers the log density of the GRF under various noise settings, with a relative scatter of $\sim6$ due to simplified integration scheme, insufficient noise realizations in Equation \ref{eqn:sk estimator} and insufficient steps.
The residual scatter shown in right panel of Figure \ref{fig:grf} are almost independent of the noise level, indicating that OTFM delivers stable accuracy across datasets with different signal‑to‑noise ratios. 

The present evaluation employs the simplest Euler integrator with 1000 time‑steps.  
Replacing it with higher‑order schemes (e.g.\ Runge–Kutta) or increasing the step count should further tighten the agreement between OTFM and the analytic benchmark. 

{While we do not explicitly test OoD detection here, the accurate density estimation suggests OTFM's potential for identifying OoD samples in such GRF datasets.} {We also expect CNN and ST statistics to achieve good performance on this simple dataset. The former has been shown to extract the full information contents on GRFs and lognormal fields \cite{lucas2021lossless, lanzieri2025optimal}, while the first-order ST coefficients are strongly correlated with the power spectrum \cite{Cheng_2020}.}

\subsection{Detecting BCM maps as OoD}
\label{sec:bcm}
\begin{figure}
    \centering
    \includegraphics[width=\linewidth]{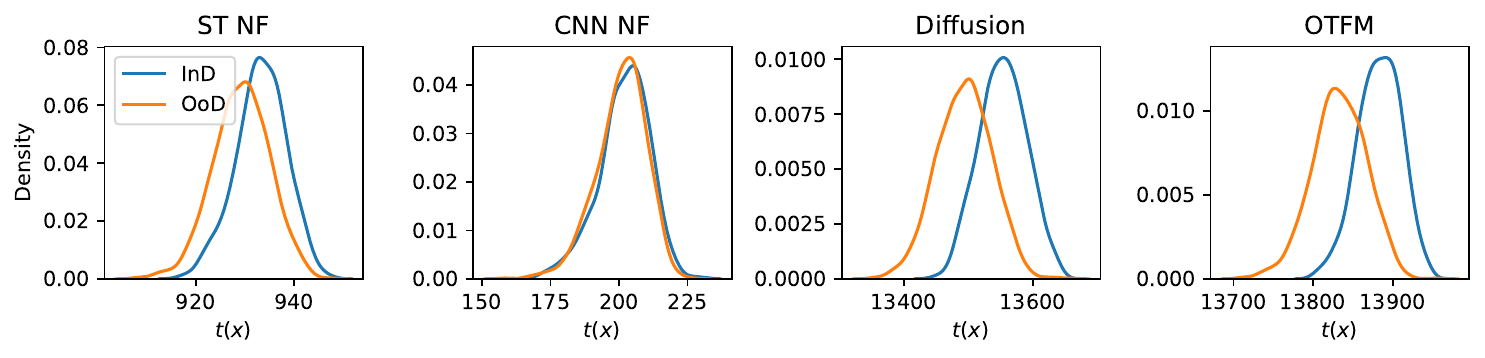}
    \caption{Example probability distribution of test statistics $t(\bm x) = \log p(x)$ of InD and OoD test dataset with noise level $n_g=30$. From left to right are $t(\bm x)$ obtained with different methods.}
    \label{fig:kde_ng30}
\end{figure}

In this subsection, we present main results from our test problem, where we try to detect the unmodeled baryonic effects when DMO simulations are available. We test the performance of each methods mentioned above on our fiducial cosmology $\theta_{\rm fid} = (\sigma_{8,\rm fid},\Omega_{m,\rm fid}) = (0.82,0.265)$. 
For feature-level detection methods like CNN and ST, the InD samples consist of 304 maps of DMO maps at fiducial cosmology, while OoD samples are 1024 BCM maps with different baryon parameters at the same fiducial cosmology. 
{This setting corresponds to an idealized case where we can perfectly infer the cosmological parameters, i.e. $p(\theta|\bm{x}_{\rm obs}) = \delta(\theta-\theta_{\rm fid})$ in Equation \ref{eqn:ppd}, to avoid sampling the full posterior which is computationlly expensive. Using the true $p(\theta|\bm{x}_{\rm obs})$ will introduce $\mathcal{O}(1)$ scatter to the test statistics and slightly degraded OoD performance. Thus our reported CNN and ST performance here represent the upper limit.}
These maps have a physical resolution of $128^2$ pixels.
For each methods mentioned before, we compute the probability density of these maps as the $t(\bm{x})$ {using Equation \ref{eqn:int}}, because the density is the most natural OoD detector. Examples with noise level $n_g=30$ are shown in Figure \ref{fig:kde_ng30}, and a well-behaved OoD detector is expected to generate clearly separable density distributions for InD and OoD sets.

For field-level detection methods like diffusion model and OTFM model, we train the model unconditionally, as the total number of cosmologies is limited making it difficult to capture the dependency of cosmologies accurately. The output log density $\log p(\bm{x})$ is thus an average 
of the likelihood over the cosmology prior and can thus be viewed as the Bayesian evidence of the observation. For higher computational efficiency, we limited the input and output dimensions to $64^2$ pixels, and the density of a $128^2$ map is the average of 4 $64^2$ map density cut from the original map.


The metric used to evaluate the performance is Receiver Operating Characteristic (ROC) curve, which is a graphical tool used in classification tasks to illustrate the performance of a binary classifier across different decision thresholds. It plots the True Positive Rate (TPR) against the FPR for different classification thresholds, effectively showing the trade-off between correctly identifying positive instances and incorrectly flagging negative ones.

The Area Under the ROC Curve (AUROC) summarizes this plot into a single number ranging from 0 to 1. A larger AUROC indicates better overall classifier performance, with 0.5 representing a random guess and 1.0 indicating a perfect model. This metric is particularly useful for comparing different {classifiers}, as it remains invariant to class distribution and threshold selection.

For feature-level detection methods like CNN and ST, we adopt conditional RealNVP to estimate their conditional log density $\log p(\bm{y}|\theta)$ with compressed features $\bm{y}$. The conditional probability performs better because of the additional information $\theta$, which is confirmed by our tests: the 
conditional density at the best 
fit value of $\theta$ gives better 
AUROC than the average over $\theta$, 
 but the difference is small and we 
 will ignore it in the following. We first compute the feature vectors from the $128^2$ maps, and use these vectors as the input of the NF and the true cosmology as the condition $\theta$.

\begin{figure}
    \centering
    \includegraphics[width=0.32\linewidth]{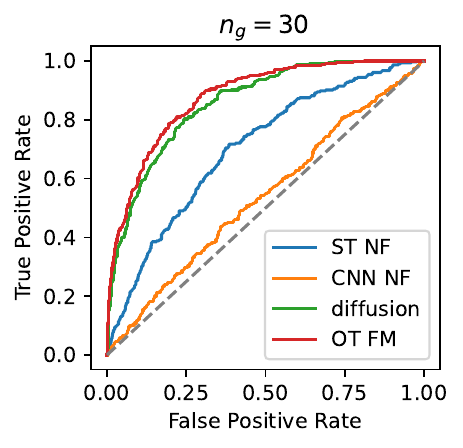}
    \includegraphics[width=0.32\linewidth]{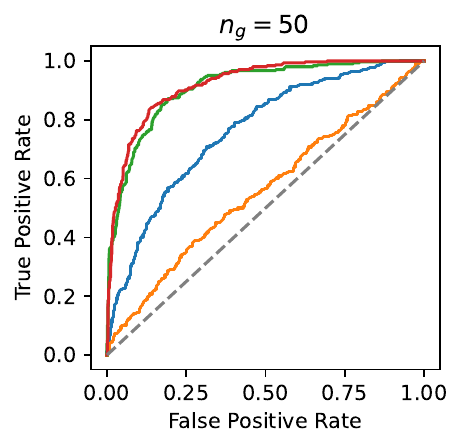}
    \includegraphics[width=0.32\linewidth]{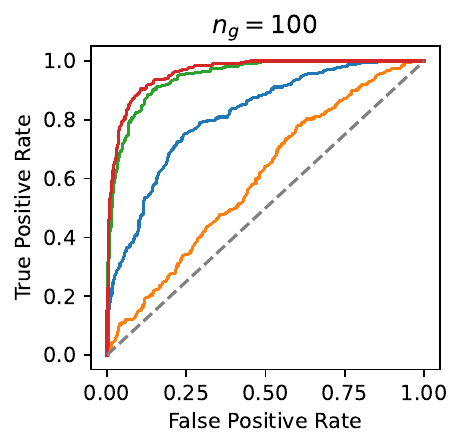}
    \caption{ROC curve of using different $\log p(\bm x)$ as OoD detection test statistics with the fiducial cosmology and resolution $128^2$.}
    \label{fig:roc}
\end{figure}

\begin{table}
\centering
\caption{OoD Detection AUROC of using different $\log p(\bm x)$ as OoD detection test statistics with the fiducial cosmology and resolution $128^2$.}
\label{tab:AUROC_ng30}
\begin{tabular}{lccc}
\toprule
 Method &  $n_g=30$ & $n_g=50$& $n_g=100$\\
\midrule
         Diffusion  &  0.85& 0.90&0.94\\
         OT FlowMatching  &  0.87& 0.92&0.95\\
         MultiscaleFlow\citep{2024PNAS..12109624D}
         \tablefootnote{In this work, the InD dataset is drawn from DMO maps at the Max-A-Priori (MAP) cosmology of the OoD maps, which gives a higher $p(\bm x|\theta)$ than the true cosmology in our case for OoD maps. However, the difference in $p(\bm x|\theta)$ between MAP and true cosmology InD maps is an order of magnitude smaller than the difference between InD and OoD $p(\bm x|\theta)$. As a result, the AUROC obtained with true cosmology InD samples would be slightly higher but not significantly so.} &  $\gtrsim0.65$& -- & --\\
         CNN  &  0.54& 0.55&0.60\\
         ST coefficient  &  0.73 &0.77 &0.81\\

\bottomrule
\end{tabular}
\end{table}

The ROC curve of different methods is shown in Figure \ref{fig:roc}, and the corresponding AUROC is presented in Table \ref{tab:AUROC_ng30}. The field-level detections significantly outperform feature-level detection, even if extra cosmology information is provided to feature-level methods. We further compared our field-level results to \citep{2024PNAS..12109624D}, concluding that our CTFM field-level approaches is much better than field-level NFs. For two CTFM field-level methods, OTFM is slightly better than diffusion as expected, for its regularized temporal evolution helps stabilize the neural network.

In the feature-level analysis, we found that ST coefficients significantly outperforms CNN, even if they have similar dimensions. 
Given that CNN is optimized by maximizing the cosmological information in the compressed features, it focuses on cosmology-dependent features and ignores other information which may be important for detecting model misspecification. This explains why CNN leads to more constraining power on cosmological parameters than ST coefficients in previous study \citep{2024JCAP...08..010S}, yet it performs poorly in OoD detection here in our experiment. 
Note that while CNN learned statistics are not able to distinguish InD data and OoD data effectively, it does not mean that CNN analysis is robust to modeling bias. Its inference can still be biased by model misspecification, since the latter can be degenerate with cosmological information. 

{When the type of OoD is known (i.e., OoD data is given at training stage), one can use supervised training techniques to train CNNs and obtain better performance \citep[e.g.][]{2018ApJ...856...68P,2020MNRAS.499..379V}. However, in many applications one does not know the type of OoD, and instead tries to search for unknown unknowns. This is the main focus of our paper, and supervised training is not applicable here since no OoD training data is given. In this scenario, we expect training the CNN with an AutoEncoder (AE) loss could improve OoD detection compared to the VMIM loss we tried, as the AE's reconstruction task necessitates learning a comprehensive representation of the input field, rather than just cosmology-specific features, and we plan to explore this in our future works.}

\subsection{Performance at different cosmologies}
\label{sec:consis}
\begin{table}[]
    \centering
    \caption{OTFM AUROC at different cosmologies}
    \begin{tabular}{c|ccc}
    \hline
       $(\sigma_8,\Omega_m)$  & $n_g=30$&$n_g=50$&$n_g=100$ \\
    \hline
       (0.82,0.268)(default)  & 0.87& 0.92&0.95\\
    \hline
        (0.717, 0.315) & 0.93& 0.96 & 0.98\\
        (0.766, 0.275) & 0.88& 0.93 & 0.96\\
        (0.768, 0.264) & 0.87& 0.92 & 0.94\\
        (0.875, 0.259) & 0.86& 0.90 & 0.93\\
        (0.864, 0.234) & 0.83& 0.87 & 0.90\\
        (0.842, 0.217) & 0.79& 0.85 & 0.87\\
    \hline

    \end{tabular}
    
    \label{tab:multicos}
\end{table}

To confirm the robustness of our method across different cosmologies, we tested the OTFM density on six additional cosmologies, with their $(\sigma_8,\Omega_m)$ values and results presented in Table~\ref{tab:multicos}. The InD samples and OoD samples are constructed in the same way as in Section \ref{sec:res}, except the cosmology is different.
By examining the AUROC values, we find that OTFM achieves consistently high performance for cosmologies that are degenerate with the default cosmology (first four entries in Table~\ref{tab:multicos}). However, in the last three additional cosmologies, where the amplitude of fluctuations decreases, the detection accuracy declines. This is likely due to smaller-scale signals being increasingly obscured by Gaussian noise, diminishing the contrast between InD and OoD samples.

\label{sec:dis}
\subsection{OoD  as a model selection}
\label{sec:sel}
\begin{figure}
    \centering
    \includegraphics[width=\linewidth]{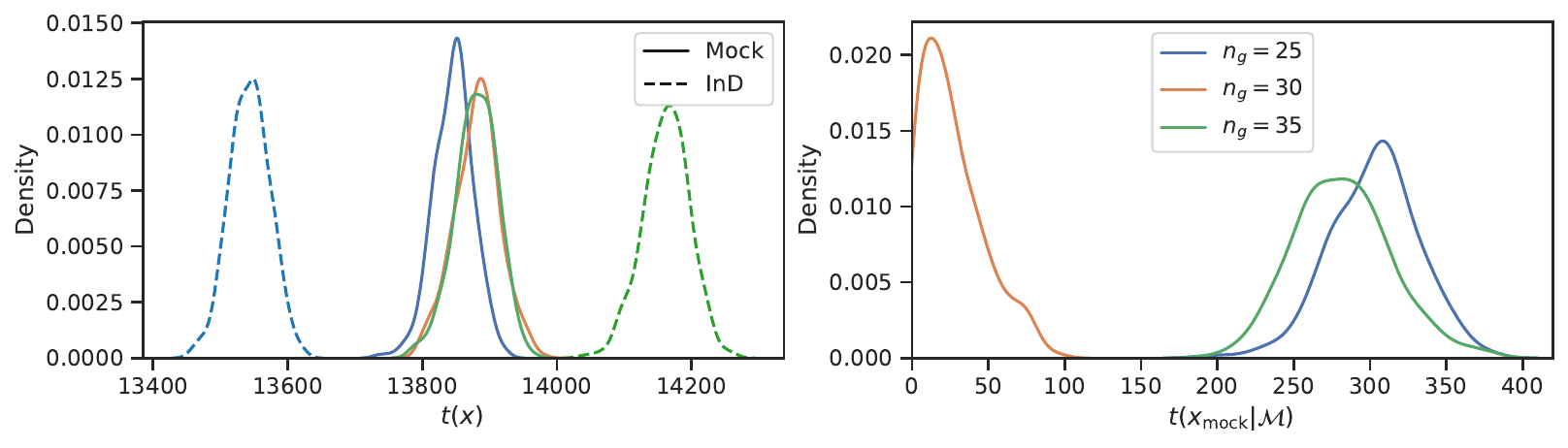}
    \caption{\textbf{Left}: the log density of the $n_{g,{\rm true}}=30$ mock observation with different models are shown in solid line, and the dashed line represents the log density of InD samples for each model, different colors represents different models. \textbf{Right}: The test statistics distribution of $n_{g,{\rm true}}=30$ samples with different model $\mathcal{M}$. 
    }
    \label{fig:model_select}
\end{figure}

A density-based OoD detector measures how much the density of an observation deviates from the typical value in the training set specified by a given model. Likewise, when we evaluate the deviation from multiple models using the same detection methods, this deviation serves as a metric of how well each model fits the observation.
A typical example would be having access to multiple 
simulations that disagree with each other \cite{2021arXiv210909747V}. In this case we may use density 
estimation as the model selection, 
choosing the simulation that 
gives the highest density of the data. 

To validate the performance of generative model likelihoods as model selectors, we test the OTFM model in a noise miscalibration scenario, as OTFM is the best test statistics according to the results in Section \ref{sec:res}. Specifically, we consider the mock observation \(\bm{x}_{\rm mock}\), which consists of 304 DMO maps with shape noise \(n_{g,{\rm {true}}}=30 {\rm arcmin^{-2}}\) at fiducial cosmology. Our three candidate models are DMO simulations with \(n_g=\{25,30,35\} {\rm arcmin^{-2}}\), and their final noise standard deviations are \(\sigma_n \approx \{0.0345, 0.0315, 0.0291\}\). We train an OTFM model for each candidate model \(\mathcal{M}\) to serve as the density estimator \(\log p(\bm{x}|\mathcal{M})\). 

We first calculated the log density of \(\bm{x}_{\rm mock}\) with different models and find their the log density distribution overlaps with each other as is shown in the left panel of Figure \ref{fig:model_select}, making it hard to identify the best model. However, the log density distribution of \(\bm{x}_{\rm mock}\) for mis-specified models significantly differ from that of InD samples, shed a light to model comparison with the difference to the typical values of the InD samples. 
In this case, we can select the best model by the false positive rate of identify them as OoD, or equivalently the log density deviation from the log density of typical set samples. Therefore, the selection metric is then defined as
\begin{equation}
    t(\bm{x}_{\rm mock}|\mathcal{M}) = |\log p(\bm{x}_{\rm mock}|\mathcal{M}) \;-\; \mathbb{E}_{\bm{x}\sim p_{\rm InD}(\bm{x})}\bigl(\log p(\bm{x}|\mathcal{M})\bigr)|,
\end{equation}
where $p_{\rm InD}(x)$ is approximated by a dataset of model output, {$\mathbb{E}_{\bm{x}\sim p_{\rm InD}(\bm{x})}\bigl(\log p(\bm{x}|\mathcal{M})\bigr)$ is the expectation of log density over the InD distribution, and is estimated empirically by taking average log density of the InD samples}. Intuitively, \(t\) grows if (i) the model is overly broad—so its typical‑set likelihood is lower than that of the \(\bm{x}_{\text{mock}}\)—or (ii) the model assigns substantially lower likelihood to \(\bm{x}_{\text{mock}}\) than to its own typical samples.
The right panel of Figure \ref{fig:model_select} shows the distribution of \(t(\bm{x}_{\rm mock}|\mathcal{M})\) for different \(n_g\) models, demonstrating that the correct \(n_g\) model attains a significantly smaller \(t(\bm{x}|\mathcal{M})\). 
With this metric OTFM successfully selects the correct model for all 304 $ \bm{x}_{\rm mock}$ maps. 



\subsection{Impact of resolution and survey area}
\label{sec:scale}

\begin{figure}
    \centering
    \includegraphics[width=0.6\linewidth]{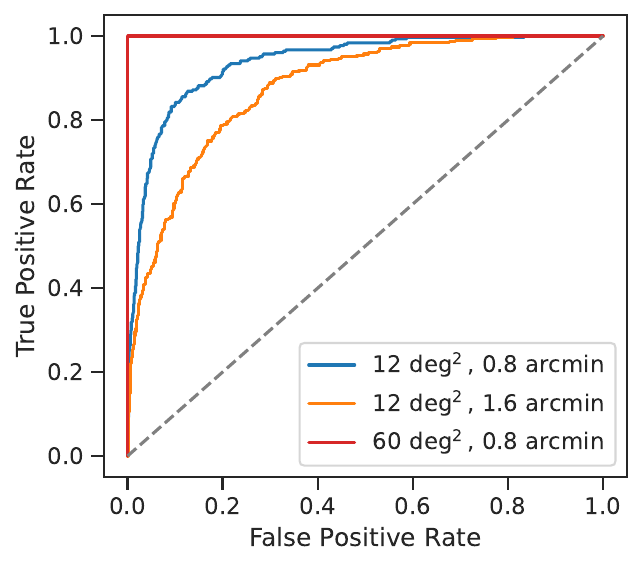}
    
    \caption{ROC curve of OTFM density with different field-of-view and resolution, as is shown in the legend. 
    }
    \label{fig:hires}
\end{figure}

Our experiments in section \ref{sec:res} is performed on mock weak lensing maps with small field-of-view ($3.5 \mathrm{deg}\times 3.5 \mathrm{deg}$) and fixed resolution (pixel size $1.64$ arcmin). This map area is significantly smaller compared to current and upcoming weak lensing surveys which normally span thousands of $\mathrm{deg}^2$, and the resolution is also much lower compared to some of the current WL analysis that aims to study baryonic effect \citep{terasawa2025exploring}. We expect our model performance to improve significantly with higher resolution \cite{2024PNAS..12109624D} and larger area, and we explicitly verify this in this section.

In Figure \ref{fig:hires} we show the OoD performance using OTFM field-level likelihood with different map area and resolutions. Here we follow the same OoD and InD setups as in Section \ref{sec:res}, but we divide the map to subfields with $64^2$ pixels, and model the likelihood of each subfields independently. The total likelihood of the map is approximated as the product of the likelihood of all subfields, ignoring the correlation between different subfields. When testing the method on larger area (60~$\deg^2$), we randomly sample 5 maps of size $3.5 \mathrm{deg}\times 3.5 \mathrm{deg}$ and combine their likelihood. We see from figure \ref{fig:hires} that increasing the resolution improves the performance of detecting baryonic physics, with AUROC increases from 0.87 to 0.94. Furthermore, the OTFM model was able to perfectly identify all OoD samples when the map size is increased to 60~$\deg^2$.

\subsection{Normalizing flows v.s. Continuous‐time Flow model}
\label{sec:badnf}
\begin{figure}
    \centering
    \includegraphics[width=0.9\linewidth]{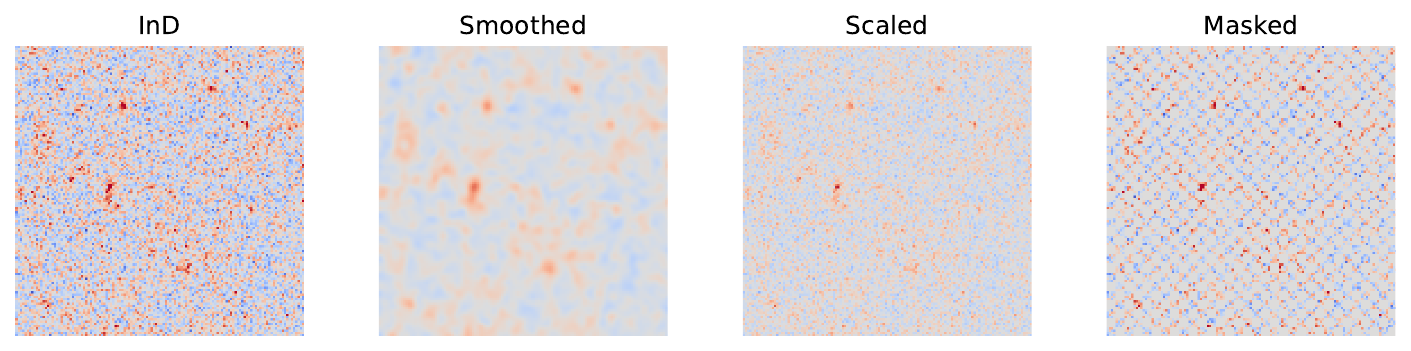}
    
    \caption{Samples of InD set, smoothed set, scaled set and masked set, respectively.}
    \label{fig:badnfsample}
\end{figure}

\begin{figure}
    \centering
    \includegraphics[width=0.9\linewidth]{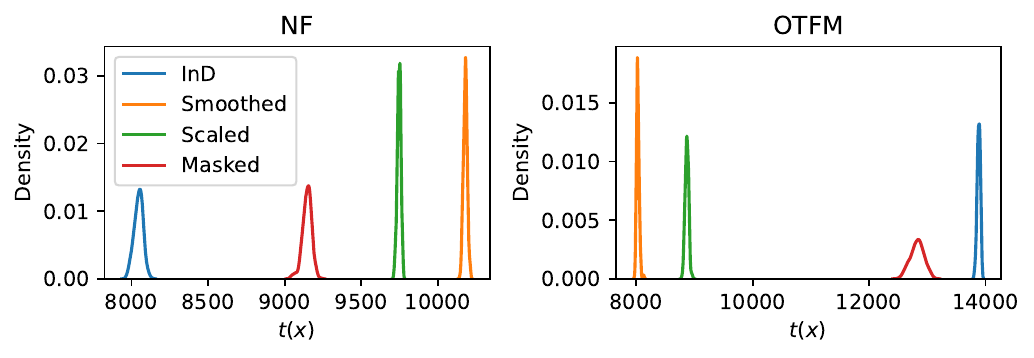}
    
    \caption{Log probability density of field-level normalizing flow and OTFM model for InD samples and different kind of OoD samples.}
    \label{fig:badnf}
\end{figure}

Although we employed CTFM in this work, NF models with certain architectures, such as GLOW \citep{2018arXiv180703039K}, are also capable of scaling to high-dimensional field-level density estimation. However, it has been observed that these NF models exhibit a preference for specific spatial structures, introducing systematic bias in the density estimation \citep{2020arXiv200608545K}. This bias renders NF suboptimal for OoD detection tasks. In our study, we confirmed this conclusion on our dataset and compared the performance of OTFM on the same test cases.

The InD samples consist of 304 DMO WL maps of size $128^2$ generated under the fiducial cosmology. We then constructed three distinct types of strong spatial biases as OoD samples, as shown in Figure \ref{fig:badnfsample}:
\begin{itemize}
    \item \textit{Smoothing}: We smooth the field using a 2D Gaussian kernel with a width of (1.5 pix, 1.5 pix) along the (x,y) directions.
    \item \textit{Scaling}: We multiply the field values by a factor of 0.5.
    \item \textit{Masking}: We mask the field using a chessboard pattern. Specifically, we divide the field into $32\times32$ blocks, each of size $4\times4$ pixels, and evenly mask half of these blocks.
\end{itemize}
Each of these OoD sets contains 304 samples. We then employed both the GLOW and OTFM models to estimate the probability density as test statistics $t(x)$; the corresponding PDFs are shown in Figure \ref{fig:badnf}. {An ideal likelihood estimator is supposed to assign lower likelihood to these artificial OoD maps, however} it is evident that GLOW misestimates the OoD sets {by assigning higher likelihood for OoD samples, as is shown in the left panel of Figure \ref{fig:badnf}. Meanwhile}, OTFM remains robust and successfully identifies all OoD samples {by assigning low likelihoods for these samples}. This limitation originates from the inductive bias inherent in the GLOW architecture, and the loss is less important than the architecture \citep{2020arXiv200608545K}. Although alternative test statistics, such as the 2-norm of the score function $\lVert \partial\log p(\bm{x})/\partial \bm{x} \rVert$ which serves as a metric for typicality, can mitigate this issue \citep{2019arXiv191203263G,2019arXiv190602994N}, they still suffer from certain biases \citep{2024arXiv240903043V}. 
In conclusion, our findings confirm that CTFM generally outperforms NF, as it has less inductive biases from the model architecture, thus, the model can learn more information from the loss function without being limited by the model inductive bias. 

\section{Conclusion}
\label{sec:con}
Future cosmological surveys are poised to deliver high-quality, high-dimensional observables, and ML-based SBI has demonstrated remarkable efficacy in extracting information from complex, field-level data. Nonetheless, the fidelity of forward modeling remains pivotal for accurate posterior estimation. In this work, we reframe the validation of forward models as an OoD detection problem—a null test to determine whether an observation originates from the distribution produced by the forward model. Among the various summary statistics examined, the field-level probability density estimated via CTFM achieves the best performance.

Our test set comprises DMO WL maps and BCM WL maps generated under identical cosmologies. We evaluate the probability density using test statistics computed at different levels and by various methods. At the feature level, we employ ST and CNN as feature extractors and train NFs separately to assess the density. At the field level, we utilize two variants of CTFM—diffusion models and OTFM—as density estimators. Notably, CTFM significantly outperforms the feature-level approaches as well as the NF-based field-level density estimator MultiscaleFlow \citep{2024PNAS..12109624D}, as quantified by AUROC. This suggests 
that there is a lot of information 
at the field level that is lost 
if one compresses the data into 
$O(100)$ features. 

Moreover, our findings indicate that the magnitude of deviation from typical models not only serves as a null test for consistency with forward modeling but also functions as a metric for model selection. 
We further demonstrate that increasing the map resolution and map size significantly enhances the OoD performance -- when the survey area reaches 60$\deg^2$ we achieves a perfect characterization of our test OoD samples with $n_g=30$ shape noise.
We also confirm that CTFM maintains consistent performance across different cosmologies. Finally, our results corroborate previous findings regarding the inductive bias of NF toward certain spatial patterns, while also revealing that CTFM is more robust against biases stemming from model architecture.

As a proof of concept, this study employs simplified forward modeling and a limited OoD test set. Future work could address these limitations by incorporating more realistic baryon models and observational systematics. 
Additionally, this work detects the modeling bias without addressing explicitly the interpretability of the results. Future work can focus on various 
splits of the data to identify 
the regions of strongest OoD detection, such as splits by scale, density etc. 

\acknowledgments


KD is supported by the National SKA Program of China (grant No.~2020SKA0110401) and NSFC (grant No.~11821303). BD acknowledges support from the Ambrose Monell Foundation, the Corning Glass Works Foundation Fellowship Fund, and the Institute for Advanced Study. This work is also supported by U.S. Department of Energy, Office of Science, Office of Advanced Scientific Computing Research under Contract No. DE-AC02-05CH11231 at Lawrence Berkeley National Laboratory to enable research for Data-intensive Machine Learning and Analysis. KD thanks Tao Jing, Ce Sui, Bhuvnesh Jain, Matias Zaldarriaga, Richard Grumitt and Xiaosheng Zhao for their helpful insights and comments.


\bibliographystyle{JHEP}
\bibliography{biblio.bib}

\providecommand{\href}[2]{#2}\begingroup\raggedright\begin{thebibliography}{10}

\bibitem{2001PhR...340..291B}
M.~{Bartelmann} and P.~{Schneider}, \emph{{Weak gravitational lensing}}, \href{https://doi.org/10.1016/S0370-1573(00)00082-X}{\emph{\physrep} {\bfseries 340} (2001) 291} [\href{https://arxiv.org/abs/astro-ph/9912508}{{\ttfamily astro-ph/9912508}}].

\bibitem{2015RPPh...78h6901K}
M.~{Kilbinger}, \emph{{Cosmology with cosmic shear observations: a review}}, \href{https://doi.org/10.1088/0034-4885/78/8/086901}{\emph{Reports on Progress in Physics} {\bfseries 78} (2015) 086901} [\href{https://arxiv.org/abs/1411.0115}{{\ttfamily 1411.0115}}].

\bibitem{2017MNRAS.465.1454H}
H.~{Hildebrandt}, M.~{Viola}, C.~{Heymans}, S.~{Joudaki}, K.~{Kuijken}, C.~{Blake} et~al., \emph{{KiDS-450: cosmological parameter constraints from tomographic weak gravitational lensing}}, \href{https://doi.org/10.1093/mnras/stw2805}{\emph{\mnras} {\bfseries 465} (2017) 1454} [\href{https://arxiv.org/abs/1606.05338}{{\ttfamily 1606.05338}}].

\bibitem{2017MNRAS.465.2033J}
S.~{Joudaki}, C.~{Blake}, C.~{Heymans}, A.~{Choi}, J.~{Harnois-Deraps}, H.~{Hildebrandt} et~al., \emph{{CFHTLenS revisited: assessing concordance with Planck including astrophysical systematics}}, \href{https://doi.org/10.1093/mnras/stw2665}{\emph{\mnras} {\bfseries 465} (2017) 2033} [\href{https://arxiv.org/abs/1601.05786}{{\ttfamily 1601.05786}}].

\bibitem{2019PASJ...71...43H}
C.~{Hikage}, M.~{Oguri}, T.~{Hamana}, S.~{More}, R.~{Mandelbaum}, M.~{Takada} et~al., \emph{{Cosmology from cosmic shear power spectra with Subaru Hyper Suprime-Cam first-year data}}, \href{https://doi.org/10.1093/pasj/psz010}{\emph{\pasj} {\bfseries 71} (2019) 43} [\href{https://arxiv.org/abs/1809.09148}{{\ttfamily 1809.09148}}].

\bibitem{2020PASJ...72...16H}
T.~{Hamana}, M.~{Shirasaki}, S.~{Miyazaki}, C.~{Hikage}, M.~{Oguri}, S.~{More} et~al., \emph{{Cosmological constraints from cosmic shear two-point correlation functions with HSC survey first-year data}}, \href{https://doi.org/10.1093/pasj/psz138}{\emph{\pasj} {\bfseries 72} (2020) 16} [\href{https://arxiv.org/abs/1906.06041}{{\ttfamily 1906.06041}}].

\bibitem{2019ApJ...873..111I}
{\v{Z}}.~{Ivezi{\'c}}, S.M.~{Kahn}, J.A.~{Tyson}, B.~{Abel}, E.~{Acosta}, R.~{Allsman} et~al., \emph{{LSST: From Science Drivers to Reference Design and Anticipated Data Products}}, \href{https://doi.org/10.3847/1538-4357/ab042c}{\emph{\apj} {\bfseries 873} (2019) 111} [\href{https://arxiv.org/abs/0805.2366}{{\ttfamily 0805.2366}}].

\bibitem{2022A&A...662A.112E}
{Euclid Collaboration}, R.~{Scaramella}, J.~{Amiaux}, Y.~{Mellier}, C.~{Burigana}, C.S.~{Carvalho} et~al., \emph{{Euclid preparation. I. The Euclid Wide Survey}}, \href{https://doi.org/10.1051/0004-6361/202141938}{\emph{\aap} {\bfseries 662} (2022) A112} [\href{https://arxiv.org/abs/2108.01201}{{\ttfamily 2108.01201}}].

\bibitem{2015arXiv150303757S}
D.~{Spergel}, N.~{Gehrels}, C.~{Baltay}, D.~{Bennett}, J.~{Breckinridge}, M.~{Donahue} et~al., \emph{{Wide-Field InfrarRed Survey Telescope-Astrophysics Focused Telescope Assets WFIRST-AFTA 2015 Report}}, \href{https://doi.org/10.48550/arXiv.1503.03757}{\emph{arXiv e-prints} (2015) arXiv:1503.03757} [\href{https://arxiv.org/abs/1503.03757}{{\ttfamily 1503.03757}}].

\bibitem{2017MNRAS.469.2737K}
T.D.~{Kitching}, J.~{Alsing}, A.F.~{Heavens}, R.~{Jimenez}, J.D.~{McEwen} and L.~{Verde}, \emph{{The limits of cosmic shear}}, \href{https://doi.org/10.1093/mnras/stx1039}{\emph{\mnras} {\bfseries 469} (2017) 2737} [\href{https://arxiv.org/abs/1611.04954}{{\ttfamily 1611.04954}}].

\bibitem{2003MNRAS.344..857T}
M.~{Takada} and B.~{Jain}, \emph{{Three-point correlations in weak lensing surveys: model predictions and applications}}, \href{https://doi.org/10.1046/j.1365-8711.2003.06868.x}{\emph{\mnras} {\bfseries 344} (2003) 857} [\href{https://arxiv.org/abs/astro-ph/0304034}{{\ttfamily astro-ph/0304034}}].

\bibitem{2003ApJ...584..559Z}
M.~{Zaldarriaga} and R.~{Scoccimarro}, \emph{{Higher Order Moments of the Cosmic Shear and Other Spin-2 Fields}}, \href{https://doi.org/10.1086/345789}{\emph{\apj} {\bfseries 584} (2003) 559} [\href{https://arxiv.org/abs/astro-ph/0208075}{{\ttfamily astro-ph/0208075}}].

\bibitem{2014MNRAS.441.2725F}
L.~{Fu}, M.~{Kilbinger}, T.~{Erben}, C.~{Heymans}, H.~{Hildebrandt}, H.~{Hoekstra} et~al., \emph{{CFHTLenS: cosmological constraints from a combination of cosmic shear two-point and three-point correlations}}, \href{https://doi.org/10.1093/mnras/stu754}{\emph{\mnras} {\bfseries 441} (2014) 2725} [\href{https://arxiv.org/abs/1404.5469}{{\ttfamily 1404.5469}}].

\bibitem{2011MNRAS.410..143S}
E.~{Semboloni}, T.~{Schrabback}, L.~{van Waerbeke}, S.~{Vafaei}, J.~{Hartlap} and S.~{Hilbert}, \emph{{Weak lensing from space: first cosmological constraints from three-point shear statistics}}, \href{https://doi.org/10.1111/j.1365-2966.2010.17430.x}{\emph{\mnras} {\bfseries 410} (2011) 143} [\href{https://arxiv.org/abs/1005.4941}{{\ttfamily 1005.4941}}].

\bibitem{2011ApJ...738...86C}
J.~{Carron}, \emph{{On the Incompleteness of the Moment and Correlation Function Hierarchy as Probes of the Lognormal Field}}, \href{https://doi.org/10.1088/0004-637X/738/1/86}{\emph{\apj} {\bfseries 738} (2011) 86} [\href{https://arxiv.org/abs/1105.4467}{{\ttfamily 1105.4467}}].

\bibitem{2009ApJ...698L..90N}
M.C.~{Neyrinck}, I.~{Szapudi} and A.S.~{Szalay}, \emph{{Rejuvenating the Matter Power Spectrum: Restoring Information with a Logarithmic Density Mapping}}, \href{https://doi.org/10.1088/0004-637X/698/2/L90}{\emph{\apjl} {\bfseries 698} (2009) L90} [\href{https://arxiv.org/abs/0903.4693}{{\ttfamily 0903.4693}}].

\bibitem{2016JCAP...11..057W}
M.~{White}, \emph{{A marked correlation function for constraining modified gravity models}}, \href{https://doi.org/10.1088/1475-7516/2016/11/057}{\emph{\jcap} {\bfseries 2016} (2016) 057} [\href{https://arxiv.org/abs/1609.08632}{{\ttfamily 1609.08632}}].

\bibitem{2000ApJ...530L...1J}
B.~{Jain} and L.~{Van Waerbeke}, \emph{{Statistics of Dark Matter Halos from Gravitational Lensing}}, \href{https://doi.org/10.1086/312480}{\emph{\apjl} {\bfseries 530} (2000) L1} [\href{https://arxiv.org/abs/astro-ph/9910459}{{\ttfamily astro-ph/9910459}}].

\bibitem{2010PhRvD..81d3519K}
J.M.~{Kratochvil}, Z.~{Haiman} and M.~{May}, \emph{{Probing cosmology with weak lensing peak counts}}, \href{https://doi.org/10.1103/PhysRevD.81.043519}{\emph{\prd} {\bfseries 81} (2010) 043519} [\href{https://arxiv.org/abs/0907.0486}{{\ttfamily 0907.0486}}].

\bibitem{2019BAAS...51c..40P}
A.~{Pisani}, E.~{Massara}, D.N.~{Spergel}, D.~{Alonso}, T.~{Baker}, Y.-C.~{Cai} et~al., \emph{{Cosmic voids: a novel probe to shed light on our Universe}}, \href{https://doi.org/10.48550/arXiv.1903.05161}{\emph{\baas} {\bfseries 51} (2019) 40} [\href{https://arxiv.org/abs/1903.05161}{{\ttfamily 1903.05161}}].

\bibitem{1994A&A...288..697M}
K.R.~{Mecke}, T.~{Buchert} and H.~{Wagner}, \emph{{Robust morphological measures for large-scale structure in the Universe}}, \href{https://doi.org/10.48550/arXiv.astro-ph/9312028}{\emph{\aap} {\bfseries 288} (1994) 697} [\href{https://arxiv.org/abs/astro-ph/9312028}{{\ttfamily astro-ph/9312028}}].

\bibitem{2012PhRvD..85j3513K}
J.M.~{Kratochvil}, E.A.~{Lim}, S.~{Wang}, Z.~{Haiman}, M.~{May} and K.~{Huffenberger}, \emph{{Probing cosmology with weak lensing Minkowski functionals}}, \href{https://doi.org/10.1103/PhysRevD.85.103513}{\emph{\prd} {\bfseries 85} (2012) 103513} [\href{https://arxiv.org/abs/1109.6334}{{\ttfamily 1109.6334}}].

\bibitem{Mallat_2012}
S.~Mallat, \emph{Group invariant scattering}, \href{https://doi.org/https://doi.org/10.1002/cpa.21413}{\emph{Communications on Pure and Applied Mathematics} {\bfseries 65} (2012) 1331} [\href{https://arxiv.org/abs/https://onlinelibrary.wiley.com/doi/pdf/10.1002/cpa.21413}{{\ttfamily https://onlinelibrary.wiley.com/doi/pdf/10.1002/cpa.21413}}].

\bibitem{Cheng_2021}
S.~{Cheng} and B.~{M{\'e}nard}, \emph{{How to quantify fields or textures? A guide to the scattering transform}}, {\emph{arXiv e-prints} (2021) arXiv:2112.01288} [\href{https://arxiv.org/abs/2112.01288}{{\ttfamily 2112.01288}}].

\bibitem{Cheng_2020}
S.~Cheng, Y.-S.~Ting, B.~M{\'{e} }nard and J.~Bruna, \emph{A new approach to observational cosmology using the scattering transform}, \href{https://doi.org/10.1093/mnras/staa3165}{\emph{\mnras} {\bfseries 499} (2020) 5902}.

\bibitem{2022PhRvD.105j3534V}
G.~{Valogiannis} and C.~{Dvorkin}, \emph{{Towards an optimal estimation of cosmological parameters with the wavelet scattering transform}}, \href{https://doi.org/10.1103/PhysRevD.105.103534}{\emph{\prd} {\bfseries 105} (2022) 103534} [\href{https://arxiv.org/abs/2108.07821}{{\ttfamily 2108.07821}}].

\bibitem{2020PhRvD.102j3506A}
E.~{Allys}, T.~{Marchand}, J.F.~{Cardoso}, F.~{Villaescusa-Navarro}, S.~{Ho} and S.~{Mallat}, \emph{{New interpretable statistics for large-scale structure analysis and generation}}, \href{https://doi.org/10.1103/PhysRevD.102.103506}{\emph{\prd} {\bfseries 102} (2020) 103506} [\href{https://arxiv.org/abs/2006.06298}{{\ttfamily 2006.06298}}].

\bibitem{2018PhRvD..97j3515G}
A.~{Gupta}, J.M.~{Zorrilla Matilla}, D.~{Hsu} and Z.~{Haiman}, \emph{{Non-Gaussian information from weak lensing data via deep learning}}, \href{https://doi.org/10.1103/PhysRevD.97.103515}{\emph{\prd} {\bfseries 97} (2018) 103515} [\href{https://arxiv.org/abs/1802.01212}{{\ttfamily 1802.01212}}].

\bibitem{2021JCAP...11..049M}
T.L.~{Makinen}, T.~{Charnock}, J.~{Alsing} and B.D.~{Wandelt}, \emph{{Lossless, scalable implicit likelihood inference for cosmological fields}}, \href{https://doi.org/10.1088/1475-7516/2021/11/049}{\emph{\jcap} {\bfseries 2021} (2021) 049} [\href{https://arxiv.org/abs/2107.07405}{{\ttfamily 2107.07405}}].

\bibitem{2022MNRAS.511.1518L}
T.~{Lu}, Z.~{Haiman} and J.M.~{Zorrilla Matilla}, \emph{{Simultaneously constraining cosmology and baryonic physics via deep learning from weak lensing}}, \href{https://doi.org/10.1093/mnras/stac161}{\emph{\mnras} {\bfseries 511} (2022) 1518} [\href{https://arxiv.org/abs/2109.11060}{{\ttfamily 2109.11060}}].

\bibitem{2023MNRAS.521.2050L}
T.~{Lu}, Z.~{Haiman} and X.~{Li}, \emph{{Cosmological constraints from HSC survey first-year data using deep learning}}, \href{https://doi.org/10.1093/mnras/stad686}{\emph{\mnras} {\bfseries 521} (2023) 2050} [\href{https://arxiv.org/abs/2301.01354}{{\ttfamily 2301.01354}}].

\bibitem{2018PhRvD..97h3004C}
T.~{Charnock}, G.~{Lavaux} and B.D.~{Wandelt}, \emph{{Automatic physical inference with information maximizing neural networks}}, \href{https://doi.org/10.1103/PhysRevD.97.083004}{\emph{\prd} {\bfseries 97} (2018) 083004} [\href{https://arxiv.org/abs/1802.03537}{{\ttfamily 1802.03537}}].

\bibitem{2024arXiv241007548M}
T.L.~{Makinen}, C.~{Sui}, B.D.~{Wandelt}, N.~{Porqueres} and A.~{Heavens}, \emph{{Hybrid Summary Statistics}}, \href{https://doi.org/10.48550/arXiv.2410.07548}{\emph{arXiv e-prints} (2024) arXiv:2410.07548} [\href{https://arxiv.org/abs/2410.07548}{{\ttfamily 2410.07548}}].

\bibitem{2022MNRAS.516.2363D}
B.~{Dai} and U.~{Seljak}, \emph{{Translation and rotation equivariant normalizing flow (TRENF) for optimal cosmological analysis}}, \href{https://doi.org/10.1093/mnras/stac2010}{\emph{\mnras} {\bfseries 516} (2022) 2363} [\href{https://arxiv.org/abs/2202.05282}{{\ttfamily 2202.05282}}].

\bibitem{2022ApJ...937...83H}
S.~{Hassan}, F.~{Villaescusa-Navarro}, B.~{Wandelt}, D.N.~{Spergel}, D.~{Angl{\'e}s-Alc{\'a}zar}, S.~{Genel} et~al., \emph{{HIFLOW: Generating Diverse HI Maps and Inferring Cosmology while Marginalizing over Astrophysics Using Normalizing Flows}}, \href{https://doi.org/10.3847/1538-4357/ac8b09}{\emph{\apj} {\bfseries 937} (2022) 83} [\href{https://arxiv.org/abs/2110.02983}{{\ttfamily 2110.02983}}].

\bibitem{2019MNRAS.488.1652H}
H.-J.~{Huang}, T.~{Eifler}, R.~{Mandelbaum} and S.~{Dodelson}, \emph{{Modelling baryonic physics in future weak lensing surveys}}, \href{https://doi.org/10.1093/mnras/stz1714}{\emph{\mnras} {\bfseries 488} (2019) 1652} [\href{https://arxiv.org/abs/1809.01146}{{\ttfamily 1809.01146}}].

\bibitem{2021arXiv210909747V}
F.~{Villaescusa-Navarro}, D.~{Angl{\'e}s-Alc{\'a}zar}, S.~{Genel}, D.N.~{Spergel}, Y.~{Li}, B.~{Wandelt} et~al., \emph{{Multifield Cosmology with Artificial Intelligence}}, \href{https://doi.org/10.48550/arXiv.2109.09747}{\emph{arXiv e-prints} (2021) arXiv:2109.09747} [\href{https://arxiv.org/abs/2109.09747}{{\ttfamily 2109.09747}}].

\bibitem{2024PNAS..12109624D}
B.~{Dai} and U.~{Seljak}, \emph{{Multiscale Flow for robust and optimal cosmological analysis}}, \href{https://doi.org/10.1073/pnas.2309624121}{\emph{Proceedings of the National Academy of Science} {\bfseries 121} (2024) e2309624121}.

\bibitem{asgari2021kids}
M.~Asgari, C.-A.~Lin, B.~Joachimi, B.~Giblin, C.~Heymans, H.~Hildebrandt et~al., \emph{Kids-1000 cosmology: Cosmic shear constraints and comparison between two point statistics}, {\emph{Astronomy \& Astrophysics} {\bfseries 645} (2021) A104}.

\bibitem{secco2022dark}
L.F.~Secco, S.~Samuroff, E.~Krause, B.~Jain, J.~Blazek, M.~Raveri et~al., \emph{Dark energy survey year 3 results: Cosmology from cosmic shear and robustness to modeling uncertainty}, {\emph{Physical Review D} {\bfseries 105} (2022) 023515}.

\bibitem{li2023hyper}
X.~Li, T.~Zhang, S.~Sugiyama, R.~Dalal, R.~Terasawa, M.M.~Rau et~al., \emph{Hyper suprime-cam year 3 results: Cosmology from cosmic shear two-point correlation functions}, {\emph{Physical Review D} {\bfseries 108} (2023) 123518}.

\bibitem{S-Dickstein2015}
J.~Sohl{-}Dickstein, E.A.~Weiss, N.~Maheswaranathan and S.~Ganguli, \emph{Deep unsupervised learning using nonequilibrium thermodynamics},  in \emph{Proceedings of the 32nd International Conference on Machine Learning, {ICML} 2015, Lille, France, 6-11 July 2015}, F.R.~Bach and D.M.~Blei, eds., vol.~37 of \emph{{JMLR} Workshop and Conference Proceedings}, pp.~2256--2265, JMLR.org, 2015, \href{http://proceedings.mlr.press/v37/sohl-dickstein15.html}{http://proceedings.mlr.press/v37/sohl-dickstein15.html}.

\bibitem{Song2021}
Y.~Song, J.~Sohl{-}Dickstein, D.P.~Kingma, A.~Kumar, S.~Ermon and B.~Poole, \emph{Score-based generative modeling through stochastic differential equations},  in \emph{9th International Conference on Learning Representations, {ICLR} 2021, Virtual Event, Austria, May 3-7, 2021}, OpenReview.net, 2021, \href{https://openreview.net/forum?id=PxTIG12RRHS}{https://openreview.net/forum?id=PxTIG12RRHS}.

\bibitem{Ho2020}
J.~Ho, A.~Jain and P.~Abbeel, \emph{Denoising diffusion probabilistic models},  in \emph{Advances in Neural Information Processing Systems 33: Annual Conference on Neural Information Processing Systems 2020, NeurIPS 2020, December 6-12, 2020, virtual}, H.~Larochelle, M.~Ranzato, R.~Hadsell, M.~Balcan and H.~Lin, eds., 2020, \href{https://proceedings.neurips.cc/paper/2020/hash/4c5bcfec8584af0d967f1ab10179ca4b-Abstract.html}{https://proceedings.neurips.cc/paper/2020/hash/4c5bcfec8584af0d967f1ab10179ca4b-Abstract.html}.

\bibitem{Lipman2022}
Y.~Lipman, R.T.Q.~Chen, H.~Ben{-}Hamu, M.~Nickel and M.~Le, \emph{Flow matching for generative modeling},  in \emph{The Eleventh International Conference on Learning Representations, {ICLR} 2023, Kigali, Rwanda, May 1-5, 2023}, OpenReview.net, 2023, \href{https://openreview.net/forum?id=PqvMRDCJT9t}{https://openreview.net/forum?id=PqvMRDCJT9t}.

\bibitem{2023MNRAS.526.1699Z}
X.~{Zhao}, Y.-S.~{Ting}, K.~{Diao} and Y.~{Mao}, \emph{{Can diffusion model conditionally generate astrophysical images?}}, \href{https://doi.org/10.1093/mnras/stad2778}{\emph{\mnras} {\bfseries 526} (2023) 1699} [\href{https://arxiv.org/abs/2307.09568}{{\ttfamily 2307.09568}}].

\bibitem{2024OJAp....7E.104S}
A.~{Schanz}, F.~{List} and O.~{Hahn}, \emph{{Stochastic Super-resolution of Cosmological Simulations with Denoising Diffusion Models}}, \href{https://doi.org/10.33232/001c.125902}{\emph{The Open Journal of Astrophysics} {\bfseries 7} (2024) 104} [\href{https://arxiv.org/abs/2310.06929}{{\ttfamily 2310.06929}}].

\bibitem{2025arXiv250204158B}
S.S.~{Boruah}, M.~{Jacob} and B.~{Jain}, \emph{{Diffusion-based mass map reconstruction from weak lensing data}}, {\emph{arXiv e-prints} (2025) arXiv:2502.04158} [\href{https://arxiv.org/abs/2502.04158}{{\ttfamily 2502.04158}}].

\bibitem{2024arXiv240717667B}
G.M.~{Barco}, A.~{Adam}, C.~{Stone}, Y.~{Hezaveh} and L.~{Perreault-Levasseur}, \emph{{Tackling the Problem of Distributional Shifts: Correcting Misspecified, High-Dimensional Data-Driven Priors for Inverse Problems}}, \href{https://doi.org/10.48550/arXiv.2407.17667}{\emph{arXiv e-prints} (2024) arXiv:2407.17667} [\href{https://arxiv.org/abs/2407.17667}{{\ttfamily 2407.17667}}].

\bibitem{2025ApJ...978...64M}
N.~{Mudur}, C.~{Cuesta-Lazaro} and D.P.~{Finkbeiner}, \emph{{Diffusion-HMC: Parameter Inference with Diffusion-model-driven Hamiltonian Monte Carlo}}, \href{https://doi.org/10.3847/1538-4357/ad8bc3}{\emph{\apj} {\bfseries 978} (2025) 64} [\href{https://arxiv.org/abs/2405.05255}{{\ttfamily 2405.05255}}].

\bibitem{gelman1996posterior}
A.~Gelman, X.-L.~Meng and H.~Stern, \emph{Posterior predictive assessment of model fitness via realized discrepancies}, {\emph{Statistica sinica} (1996) 733}.

\bibitem{VMIM}
D.~Barber and F.~Agakov, \emph{The im algorithm: a variational approach to information maximization},  in \emph{Proceedings of the 17th International Conference on Neural Information Processing Systems}, NIPS'03, (Cambridge, MA, USA), p.~201–208, MIT Press, 2003.

\bibitem{lucas2021lossless}
M.T.~Lucas, C.~Tom, A.~Justin et~al., \emph{Lossless, scalable implicit likelihood inference for cosmological fields}, {\emph{JCAP} {\bfseries 11} (2021) }.

\bibitem{lanzieri2025optimal}
D.~Lanzieri, J.~Zeghal, T.L.~Makinen, A.~Boucaud, J.-L.~Starck and F.~Lanusse, \emph{Optimal neural summarization for full-field weak lensing cosmological implicit inference}, {\emph{Astronomy \& Astrophysics} {\bfseries 697} (2025) A162}.

\bibitem{2024JCAP...08..010S}
D.~{Sharma}, B.~{Dai} and U.~{Seljak}, \emph{{A comparative study of cosmological constraints from weak lensing using Convolutional Neural Networks}}, \href{https://doi.org/10.1088/1475-7516/2024/08/010}{\emph{\jcap} {\bfseries 2024} (2024) 010} [\href{https://arxiv.org/abs/2403.03490}{{\ttfamily 2403.03490}}].

\bibitem{2016arXiv160508803D}
L.~Dinh, J.~Sohl{-}Dickstein and S.~Bengio, \emph{Density estimation using real {NVP}},  in \emph{5th International Conference on Learning Representations, {ICLR} 2017, Toulon, France, April 24-26, 2017, Conference Track Proceedings}, OpenReview.net, 2017, \href{https://openreview.net/forum?id=HkpbnH9lx}{https://openreview.net/forum?id=HkpbnH9lx}.

\bibitem{Andreux_2020}
M.~Andreux, T.~Angles, G.~Exarchakis, R.~Leonarduzzi, G.~Rochette, L.~Thiry et~al., \emph{Kymatio: Scattering transforms in python}, {\emph{Journal of Machine Learning Research} {\bfseries 21} (2020) 1}.

\bibitem{2016cvpr.confE...1H}
K.~{He}, X.~{Zhang}, S.~{Ren} and J.~{Sun}, \emph{{Deep Residual Learning for Image Recognition}},  in \emph{2016 IEEE Conference on Computer Vision and Pattern Recognition (CVPR}, p.~1, June, 2016, \href{https://doi.org/10.1109/CVPR.2016.90}{DOI} [\href{https://arxiv.org/abs/1512.03385}{{\ttfamily 1512.03385}}].

\bibitem{chen2018neural}
R.T.~Chen, Y.~Rubanova, J.~Bettencourt and D.K.~Duvenaud, \emph{Neural ordinary differential equations}, {\emph{Advances in neural information processing systems} {\bfseries 31} (2018) }.

\bibitem{Skilling1989}
J.~Skilling, \emph{The eigenvalues of mega-dimensional matrices},  in \emph{Maximum Entropy and Bayesian Methods: Cambridge, England, 1988}, J.~Skilling, ed., (Dordrecht), pp.~455--466, Springer Netherlands (1989), \href{https://doi.org/10.1007/978-94-015-7860-8_48}{DOI}.

\bibitem{hutchinson1989stochastic}
M.F.~Hutchinson, \emph{A stochastic estimator of the trace of the influence matrix for laplacian smoothing splines}, {\emph{Communications in Statistics-Simulation and Computation} {\bfseries 18} (1989) 1059}.

\bibitem{SongJ2020}
J.~Song, C.~Meng and S.~Ermon, \emph{Denoising diffusion implicit models},  in \emph{9th International Conference on Learning Representations, {ICLR} 2021, Virtual Event, Austria, May 3-7, 2021}, OpenReview.net, 2021, \href{https://openreview.net/forum?id=St1giarCHLP}{https://openreview.net/forum?id=St1giarCHLP}.

\bibitem{sarkka2019applied}
S.~S{\"a}rkk{\"a} and A.~Solin, \emph{Applied stochastic differential equations}, vol.~10, Cambridge University Press (2019).

\bibitem{mccann1997convexity}
R.J.~McCann, \emph{A convexity principle for interacting gases}, {\emph{Advances in mathematics} {\bfseries 128} (1997) 153}.

\bibitem{2023arXiv230200482T}
A.~Tong, K.~Fatras, N.~Malkin, G.~Huguet, Y.~Zhang, J.~Rector{-}Brooks et~al., \emph{Improving and generalizing flow-based generative models with minibatch optimal transport}, {\emph{Trans. Mach. Learn. Res.} {\bfseries 2024} (2024) }.

\bibitem{2005MNRAS.364.1105S}
V.~{Springel}, \emph{{The cosmological simulation code GADGET-2}}, \href{https://doi.org/10.1111/j.1365-2966.2005.09655.x}{\emph{\mnras} {\bfseries 364} (2005) 1105} [\href{https://arxiv.org/abs/astro-ph/0505010}{{\ttfamily astro-ph/0505010}}].

\bibitem{1992grle.book.....S}
P.~{Schneider}, J.~{Ehlers} and E.E.~{Falco}, \emph{{Gravitational Lenses}}, Springer (1992), \href{https://doi.org/10.1007/978-3-662-03758-4}{10.1007/978-3-662-03758-4}.

\bibitem{2021MNRAS.506.3406L}
T.~{Lu} and Z.~{Haiman}, \emph{{The impact of baryons on cosmological inference from weak lensing statistics}}, \href{https://doi.org/10.1093/mnras/stab1978}{\emph{\mnras} {\bfseries 506} (2021) 3406} [\href{https://arxiv.org/abs/2104.04165}{{\ttfamily 2104.04165}}].

\bibitem{2020MNRAS.495.4800A}
G.~{Aric{\`o}}, R.E.~{Angulo}, C.~{Hern{\'a}ndez-Monteagudo}, S.~{Contreras}, M.~{Zennaro}, M.~{Pellejero-Iba{\~n}ez} et~al., \emph{{Modelling the large-scale mass density field of the universe as a function of cosmology and baryonic physics}}, \href{https://doi.org/10.1093/mnras/staa1478}{\emph{\mnras} {\bfseries 495} (2020) 4800} [\href{https://arxiv.org/abs/1911.08471}{{\ttfamily 1911.08471}}].

\bibitem{2018ApJ...856...68P}
M.~{Pourrahmani}, H.~{Nayyeri} and A.~{Cooray}, \emph{{LensFlow: A Convolutional Neural Network in Search of Strong Gravitational Lenses}}, \href{https://doi.org/10.3847/1538-4357/aaae6a}{\emph{ApJ} {\bfseries 856} (2018) 68} [\href{https://arxiv.org/abs/1705.05857}{{\ttfamily 1705.05857}}].

\bibitem{2020MNRAS.499..379V}
A.~{Vafaei Sadr}, B.A.~{Bassett}, N.~{Oozeer}, Y.~{Fantaye} and C.~{Finlay}, \emph{{Deep learning improves identification of Radio Frequency Interference}}, \href{https://doi.org/10.1093/mnras/staa2724}{\emph{MNRAS} {\bfseries 499} (2020) 379} [\href{https://arxiv.org/abs/2005.08992}{{\ttfamily 2005.08992}}].

\bibitem{terasawa2025exploring}
R.~Terasawa, X.~Li, M.~Takada, T.~Nishimichi, S.~Tanaka, S.~Sugiyama et~al., \emph{Exploring the baryonic effect signature in the hyper suprime-cam year 3 cosmic shear two-point correlations on small scales: The s 8 tension remains present}, {\emph{Physical Review D} {\bfseries 111} (2025) 063509}.

\bibitem{2018arXiv180703039K}
D.P.~Kingma and P.~Dhariwal, \emph{Glow: Generative flow with invertible 1x1 convolutions},  in \emph{Advances in Neural Information Processing Systems 31: Annual Conference on Neural Information Processing Systems 2018, NeurIPS 2018, December 3-8, 2018, Montr{\'{e}}al, Canada}, S.~Bengio, H.M.~Wallach, H.~Larochelle, K.~Grauman, N.~Cesa{-}Bianchi and R.~Garnett, eds., pp.~10236--10245, 2018, \href{https://proceedings.neurips.cc/paper/2018/hash/d139db6a236200b21cc7f752979132d0-Abstract.html}{https://proceedings.neurips.cc/paper/2018/hash/d139db6a236200b21cc7f752979132d0-Abstract.html}.

\bibitem{2020arXiv200608545K}
P.~Kirichenko, P.~Izmailov and A.G.~Wilson, \emph{Why normalizing flows fail to detect out-of-distribution data},  in \emph{Advances in Neural Information Processing Systems 33: Annual Conference on Neural Information Processing Systems 2020, NeurIPS 2020, December 6-12, 2020, virtual}, H.~Larochelle, M.~Ranzato, R.~Hadsell, M.~Balcan and H.~Lin, eds., 2020, \href{https://proceedings.neurips.cc/paper/2020/hash/ecb9fe2fbb99c31f567e9823e884dbec-Abstract.html}{https://proceedings.neurips.cc/paper/2020/hash/ecb9fe2fbb99c31f567e9823e884dbec-Abstract.html}.

\bibitem{2019arXiv191203263G}
W.~Grathwohl, K.~Wang, J.~Jacobsen, D.~Duvenaud, M.~Norouzi and K.~Swersky, \emph{Your classifier is secretly an energy based model and you should treat it like one},  in \emph{8th International Conference on Learning Representations, {ICLR} 2020, Addis Ababa, Ethiopia, April 26-30, 2020}, OpenReview.net, 2020, \href{https://openreview.net/forum?id=Hkxzx0NtDB}{https://openreview.net/forum?id=Hkxzx0NtDB}.

\bibitem{2019arXiv190602994N}
E.~{Nalisnick}, A.~{Matsukawa}, Y.~{Whye Teh} and B.~{Lakshminarayanan}, \emph{{Detecting Out-of-Distribution Inputs to Deep Generative Models Using Typicality}}, \href{https://doi.org/10.48550/arXiv.1906.02994}{\emph{arXiv e-prints} (2019) arXiv:1906.02994} [\href{https://arxiv.org/abs/1906.02994}{{\ttfamily 1906.02994}}].

\bibitem{2024arXiv240903043V}
C.~{Viviers}, A.~{Valiuddin}, F.~{Caetano}, L.~{Abdi}, L.~{Filatova}, P.~{de With} et~al., \emph{{Can Your Generative Model Detect Out-of-Distribution Covariate Shift?}}, \href{https://doi.org/10.48550/arXiv.2409.03043}{\emph{arXiv e-prints} (2024) arXiv:2409.03043} [\href{https://arxiv.org/abs/2409.03043}{{\ttfamily 2409.03043}}].

\bibitem{2015arXiv150504597R}
O.~Ronneberger, P.~Fischer and T.~Brox, \emph{U-net: Convolutional networks for biomedical image segmentation},  in \emph{Medical Image Computing and Computer-Assisted Intervention - {MICCAI} 2015 - 18th International Conference Munich, Germany, October 5 - 9, 2015, Proceedings, Part {III}}, N.~Navab, J.~Hornegger, W.M.W.~III and A.F.~Frangi, eds., vol.~9351 of \emph{Lecture Notes in Computer Science}, pp.~234--241, Springer, 2015, \href{https://doi.org/10.1007/978-3-319-24574-4\_28}{DOI}.

\bibitem{2017arXiv170603762V}
A.~{Vaswani}, N.~{Shazeer}, N.~{Parmar}, J.~{Uszkoreit}, L.~{Jones}, A.N.~{Gomez} et~al., \emph{{Attention Is All You Need}}, \href{https://doi.org/10.48550/arXiv.1706.03762}{\emph{arXiv e-prints} (2017) arXiv:1706.03762} [\href{https://arxiv.org/abs/1706.03762}{{\ttfamily 1706.03762}}].

\bibitem{2023JOSS....8.5361S}
V.~{Stimper}, D.~{Liu}, A.~{Campbell}, V.~{Berenz}, L.~{Ryll}, B.~{Sch{\"o}lkopf} et~al., \emph{{normflows: A PyTorch Package for Normalizing Flows}}, \href{https://doi.org/10.21105/joss.05361}{\emph{The Journal of Open Source Software} {\bfseries 8} (2023) 5361} [\href{https://arxiv.org/abs/2302.12014}{{\ttfamily 2302.12014}}].

\end{thebibliography}\endgroup

\appendix
\section{Technical specifications of machine learning models employed in this work}
\label{app:model}
In this appendix we present the detailed structure of the neural networks used in this work.

\subsection{CNN feature compressor}
The CNN feature compressor used in Section \ref{sec:feature} is a ResNet \citep{2016cvpr.confE...1H} with 34 convolutional layers. Despite of input and output layers, this network consists of 16 basic residual blocks, each consists of two convolutional layers.
Residual blocks addresses the vanishing gradient problem by introducing skip connections around one convolutional layers. 
Formally, rather than learning a direct mapping $H(\bm{x})$, each residual block is designed to learn a residual function $F(\bm{x}) := H(\bm{x}) - \bm{x}$.
Hence, the block’s output becomes $\bm{y} = F(\bm{x}) + \bm{x}$.
Here each block consists of two convolutional layers, plus an identity or projection shortcut bypassing them. 

The 16 residual blocks are sequentially grouped into 4 groups with $\{3,4,6,3\}$ blocks, each group has $\{64,128,256,512\}$ channels, after each group, the map is $2\times$ downsampled. Finally, the map is average-pooled to 512 feature maps and pass through a fully-connected layer to 128 output features.

\subsection{RealNVP}
We use RealNVP \cite{2016arXiv160508803D} for density estimation of feature vectors, mentioned in Section \ref{sec:feature}. The RealNVP is a type of normalizing flow designed to model complex, high-dimensional distributions by applying an invertible, \emph{coupling-layer-based} transformation. Specifically, for an input vector \(\bm{x} = (x_{1}, x_{2}, \dots, x_{D})\), one \emph{flow} in RealNVP splits \(\bm{x}\) into two disjoint subsets (e.g., \(\bm{x}_{1:k}\) and \(\bm{x}_{k+1:D}\)), leaving one subset unchanged while transforming the other with learnable \emph{scale} and \emph{shift} functions:
\[
\tilde{\bm{x}}_{k+1:D} 
\;=\;
\bm{x}_{k+1:D}\,\odot\,\exp\bigl(s(\bm{x}_{1:k}\mid\theta)\bigr)
\;+\;
t\bigl(\bm{x}_{1:k}\mid\theta\bigr),
\]
where \(s(\cdot)\) and \(t(\cdot)\) are neural networks conditioned on the “unchanged” subset. By alternating which subset is updated across multiple coupling layers, RealNVP achieves strictly invertible transformations that enable exact likelihood computation and facilitate stable, gradient-based training. Our conditional RealNVP consists of 4 flows, and each flow uses a 3-layer multi-layer perceptron (MLP) as \(s(\cdot)\) and another 3-layer MLP as \(t(\cdot)\).

\subsection{U-Net in CTFM}

In Section \ref{sec:field}, a neural network is required for diffusion to predict the score function, while in OTFM it is used to predict the $f(\bm x_t,t)$ term in the ODE. We employ similar U-Nets \citep{2015arXiv150504597R} in both case. 

The U-Net architecture was originally designed for biomedical image segmentation and is characterized by a symmetric encoder-decoder structure with skip connections. These skip connections merge high-resolution features from the encoder with the decoder output, preserving spatial details that are crucial for generative tasks. In recent applications, such as diffusion models and FM, the standard U-Net is often augmented with attention blocks \citep{2017arXiv170603762V} to capture long-range dependencies.

Given an intermediate feature map $\mathbf{X} \in \mathbb{R}^{H \times W \times C}$, an attention block applies a self-attention mechanism to dynamically reweight the features. The process is described as follows:
\begin{enumerate}
    \item \textbf{Linear Projections:} The input features are projected into queries, keys, and values:
    \[
    \mathbf{Q} = \mathbf{X}\mathbf{W}_Q,\quad \mathbf{K} = \mathbf{X}\mathbf{W}_K,\quad \mathbf{V} = \mathbf{X}\mathbf{W}_V,
    \]
    where $\mathbf{W}_Q,\mathbf{W}_K,\mathbf{W}_V \in \mathbb{R}^{C \times d}$ are learnable weight matrices and $d$ is the dimension of the projected space.

    \item \textbf{Scaled Dot-Product Attention:} The attention weights are computed by taking the dot product of the queries and keys, scaling by $\sqrt{d}$ to ensure stable gradients, and applying the softmax function:
    \[
    \mathbf{A} = \operatorname{softmax}\!\left(\frac{\mathbf{Q}\mathbf{K}^\top}{\sqrt{d}}\right).
    \]
    
    \item \textbf{Output Computation:} Finally, the output of the attention block is obtained as a weighted sum of the values:
    \[
    \mathbf{Y} = \mathbf{A}\mathbf{V}.
    \]
\end{enumerate}
The output $\mathbf{Y}$ is merged with the original input via a residual connection and further refined with normalization and feed-forward layers. 

Apart from the input and output layers, our U-Net comprises 36 residual blocks in the encoder and an equal number in the decoder. Every three blocks, the feature map is downsampled by a factor of \(2\times\) in the encoder and upsampled by a factor of \(2\times\) in the decoder. When the map’s resolution falls below \(16^2\), the middle block within each group of three at the same resolution is augmented with the attention mechanism. In this module, four distinct sets of \(\mathbf{Q}\), \(\mathbf{K}\), and \(\mathbf{V}\) are computed, constituting a so-called four-head attention, and the four $\mathbf{Y}$s are subsequently convolved to produce the final residual function learned by the block. This U-Net configuration is consistently applied in both the diffusion model and the OTFM across varying resolutions.

\subsection{GLOW}

The GLOW model is used in Section \ref{sec:badnf} to study the impact of its inductive bias in OoD detection. GLOW \citep{2018arXiv180703039K} is a specialized architecture within the class of normalizing flows, relying on three primary invertible operations to ensure efficient training and sampling:
\begin{itemize}
  \item \textbf{Actnorm}: A per-channel affine transformation 
  \[
  \mathbf{y}_{c,i,j} = s_c\,\mathbf{x}_{c,i,j} + b_c,
  \]
  where \(s_c\) and \(b_c\) are trainable scale and bias parameters for the \(c\)-th channel. The log-Jacobian contribution is
  \[
  \log \bigl|\det(J_{\text{actnorm}})\bigr| = H\,W \sum_{c=1}^{C}\log|s_c|.
  \]

  \item \textbf{Invertible \(1 \times 1\) Convolution}: A learned weight matrix \(\mathbf{W} \in \mathbb{R}^{C\times C}\) is convolved across channels,
  \[
  \mathbf{y}_{:,i,j} = \mathbf{W}\,\mathbf{x}_{:,i,j}.
  \]
  The log-determinant for each spatial location \((i,j)\) is
  \[
  \log \bigl|\det(J_{1\times1})\bigr| = (H\,W)\,\log|\det(\mathbf{W})|.
  \]

  \item \textbf{Affine Coupling Layers}: Split the input \(\mathbf{x} = (\mathbf{x}_a,\mathbf{x}_b)\) and update one part conditioned on the other,
  \[
  \mathbf{y}_a = \mathbf{x}_a,\quad
  \mathbf{y}_b = \mathbf{x}_b \odot \exp\bigl(s(\mathbf{x}_a)\bigr) + t(\mathbf{x}_a),
  \]
  where \(s(\cdot)\) and \(t(\cdot)\) are neural networks producing per-element scale and shift terms. The log-Jacobian is 
  \[
  \log \bigl|\det(J_{\text{coupling}})\bigr| = \sum_i s(\mathbf{x}_a)_i.
  \]
\end{itemize}
Each GLOW block in our work consists of one actnorm layer, one $1\times1$ convolution layer, and one affine coupling layer.

Our GLOW model is implemented with \texttt{normflows} \citep{2023JOSS....8.5361S}\footnote{\url{https://github.com/VincentStimper/normalizing-flows}}, possessing a multi-scale nature. Let \( \bm x \) be the original \(64^2\) map. We partition \( \bm x \) into non-overlapping \(2 \times 2\) blocks, and for each block, we assign the pixel in each block in to four maps . Consequently, each submap \( \bm x_i \) is a \(32^2\) map. For the first $32^2$ map, we decompose it into 4 $16^2$ maps, and for the first $16^2$ map we decompose to 4 $8^2$ maps. For the total 10 maps with different resolutions, each of them goes though 16 GLOW blocks to the final target Gaussian distribution.


\end{document}